\documentclass[fleqn,usenatbib]{mnras}
\usepackage[T1]{fontenc}
\usepackage{color,soul}
\usepackage{ae,aecompl}
\usepackage{graphicx}% Including figure files
\usepackage{amsmath}% Advanced maths commands
\usepackage{amssymb}% Extra maths symbols
\usepackage{txfonts}

\title[sSFRs across the $M_{\rm bh}$-$M_{\rm *}$ diagram] 
{Specific star formation rates in the $M_{\rm bh}$-$M_{\rm *,sph}$ diagram 
and the evolutionary pathways of galaxies across the sSFR-$M_{\rm *}$ diagram}

\author[Graham et al.]{
Alister W.\ Graham$^{1}$\thanks{E-mail: AGraham@swin.edu.au},
T.\ H.\ Jarrett$^{2}$, 
and M.\ E.\ Cluver$^{1,3}$
\\
$^{1}$Centre for Astrophysics and Supercomputing, Swinburne University of
Technology, Hawthorn, VIC 3122, Australia\\
$^{2}$ Department of Astronomy, University of Cape Town, Rondebosch, South Africa\\
$^{3}$ Department of Physics and Astronomy, University of the Western Cape,
Robert Sobukwe Road, Bellville, 7535, South Africa
}

\date{Accepted XXX. Received YYY; in original form ZZZ}
\pubyear{2023}

\begin{document}
\label{firstpage}
\pagerange{\pageref{firstpage}--\pageref{lastpage}}
\maketitle

\begin{abstract}

It has been suggested that the bulge-to-total stellar mass ratio or feedback
from black holes (BHs), traced by the BH-to-(total stellar) mass ratio, might
establish a galaxy's specific star formation rate (sSFR).  We reveal that a
galaxy's morphology --- reflecting its formation history, particularly
accretions and mergers --- is a far better determinant of the sSFR.
Consequently, we suggest that galaxy formation models which regulate the sSFR
primarily through BH feedback prescriptions or bulge-regulated disc
fragmentation consider acquisitions and mergers which establish the galaxy
morphology.  We additionally make several new observations regarding current
($z\sim0$) star-formation rates. 
(i) Galaxies with little to no star formation have bulges with an extensive
range of stellar masses; bulge mass does not dictate presence/absence on the
`star-forming main sequence'.
(ii) The (wet merger)-built, dust-rich S0 galaxies are the `green valley'
bridging population between elliptical galaxies on the `red sequence' and
spiral galaxies on the blue star-forming main sequence.
(iii) The dust-poor S0 galaxies are not on the star-forming main sequence nor
in the `green valley'. Instead, they wait in the field for gas accretion
and/or minor mergers to transform them into spiral galaxies. 
Mid-infrared sample selection can miss these (primordial) low dust-content and
low stellar-luminosity S0 galaxies.  Finally, the appearance of the
quasi-triangular-shaped galaxy-assembly sequence, 
previously dubbed the Triangal, 
which tracks the morphological evolution of galaxies, 
is revealed in the sSFR-(stellar mass) diagram.

\end{abstract}


\section{Introduction}
\label{Sec_Intro}

Positions in the (black hole mass, $M_{\rm bh}$)-(spheroid's stellar mass,
$M_{\rm *,sph}$) diagram \citep{1995ARA&A..33..581K, 1998AJ....115.2285M,
  2000MNRAS.317..488S, 2001ApJ...553..677L} are now known to reflect a
galaxy's morphology and thus the formation processes establishing the
galaxies' morphologies.  \citet{2016ApJ...817...21S} introduced the notion of
a red and blue sequence in the $M_{\rm bh}$-$M_{\rm *,sph}$ diagram for
early-type galaxies (ETGs) and late-type galaxies (LTGs), respectively.  The
steep blue sequence for LTGs \citep{2019ApJ...873...85D, Graham:Sahu:22a} is
defined by a near-quadratic or `super-quadratic'
relation\footnote{\citet[][their footnote~14]{2015ApJ...798...54G} used the
term `super-quadratic' to denote an exponent in a scaling relation between 2
and 3.} and the ETG classification has since been considerably
refined. Indeed, the well-known near-linear `red sequence' is now recognised
as intensely dependent on the sample selection \citep{2019ApJ...876..155S,
  Graham-triangal}.

Separating the elliptical (E) galaxies from the lenticular (S0) 
galaxies\footnote{The S0 galaxies in our sample are spiralless galaxies with a
substantial, i.e., not just nuclear, disc component.} results in $M_{\rm bh}$-$M_{\rm
  *,sph}$ relations with a logarithmic slope of 1.6--1.9
\citep{2019ApJ...876..155S, Graham:Sahu:22a}.  Upon making a further
distinction between dust-poor and dust-rich S0 galaxies, a galaxy-by-galaxy
search of the literature revealed \citep[][see the references in
  table~2]{Graham-S0} that bulk of the latter population were built from
major\footnote{The term `major merger' refers to a roughly equal mass merger
event, as opposed to a `minor merger'.}  `wet' mergers, i.e., gas-rich
mergers, while the dust-poor S0 galaxies were not. Both types of S0 galaxy
follow their own near-quadratic or super-quadratic $M_{\rm bh}$-$M_{\rm
  *,sph}$ relation with the galaxy sample size sufficient to reveal
substantially different $M_{\rm bh}/M_{\rm *,sph}$ ratios at a given spheroid
stellar mass \citep{Graham-S0}.  Furthermore, the spiral (S) galaxies form a
bridging population between the dust-poor and dust-rich S0 galaxies. Moreover,
the stellar masses of these disc galaxies and their spheroids suggests that
the S galaxies originated from what are now the dust-poor S0 galaxies, likely
due to ongoing fuelling and gravitational perturbations from accretions and
minor mergers, as discussed in \citet{Graham-triangal}.

These advances are important for understanding the gravitational wave
background \citep{2020ApJ...905L..34A, 2021MNRAS.502L..99M,
  2022arXiv220405434A} because the somewhat primordial, dust-poor S0 galaxies
\citep{2013MSAIS..25...93S} are not built from major mergers involving
colliding supermassive black holes (SMBHs).  Therefore, they should be
excluded \citep[as done in][]{Graham-triangal} when establishing the $M_{\rm
  bh}$-$M_{\rm *,gal}$ relation for use in studies of gravitational waves.
Furthermore, upon distinguishing between ordinary E galaxies and brightest
cluster galaxies (BCGs), they too follow their own separate near-quadratic
$M_{\rm bh}$-$M_{\rm *,sph}$ relations with independent normalisation points,
i.e., zero-points \citep{Graham-triangal}.
The E galaxies are predominantly built from major dry mergers, evinced by the
E galaxies' depleted stellar cores \citep{1980Natur.287..307B,
  2001ApJ...563...34M, 2004ApJ...613L..33G}, the (SMBH and spheroid) mass jump
when transitioning from the S0 to the E galaxy relations in the $M_{\rm
  bh}$-$M_{\rm *,sph}$ diagram \citep[][section~3.4]{Graham:Sahu:22a,
  Graham-S0}, plus the presence of shells and multiple nuclei
\citep[e.g.][]{1986ApJ...306..110F, 1988ApJ...325...49L, 1992ApJ...399L.117H}.

Given these developments, active galactic nuclei (AGN) feedback models
producing a near-linear $M_{\rm bh}$-$M_{\rm *,gal}$ relation for
(spheroid-dominated) E galaxies appear to be neither accurate (because the E
galaxy $M_{\rm bh}$-$M_{\rm *,gal}$ relation has a near-quadratic slope) nor
appropriate \citep[because `dry' mergers, i.e., gas-poor mergers, rather than
  AGN-regulated star formation, established the E galaxies and their $M_{\rm
    bh}$-$M_{\rm *,gal}$ and $M_{\rm bh}$-$\sigma$ scaling
  relations,][]{2023MNRAS.518.6293G}.  Furthermore, for the spiral (S)
galaxies which are undergoing star formation, one observes $M_{\rm bh} \propto
M_{\rm *,gal}^3$ \citep{2018ApJ...869..113D, Graham-triangal}, which is
steeper than the S galaxy $M_{\rm bh} \propto M_{\rm *,sph}^2$ relation due to
the systematically larger ratio of galaxy-to-spheroid stellar mass in S
galaxies with smaller black holes.

Building on these (galaxy morphology)-dependent relations 
in the $M_{\rm bh}$-$M_{\rm *,sph}$ and $M_{\rm bh}$-$M_{\rm *,gal}$ diagrams, 
\citet{Graham-triangal} explains the speciation of galaxies,
i.e.,  the transformation between different galaxy types, in terms of
accretions and mergers.  In this picture, (i) S galaxies grow through accretion and
minor mergers onto what were armless disc galaxies (today's dust-poor S0
galaxies), (ii) the gas-rich major merger of S galaxies builds dust-rich
S0 galaxies\footnote{One may expect a high fraction of quasar AGN in these
  dust-rich S0 galaxies when they are built.}, and (iii) the relatively 
major dry merger of S0 galaxies with substantial spheroidal-components  
builds ellicular (ES)\footnote{\citet{1966ApJ...146...28L} introduced
  the ES galaxy notation for ETGs with an intermediate-scale disc fully 
  embedded within their spheroid.} and E galaxies.\footnote{These E galaxies
  built from relatively major dry mergers may not display a substantial AGN.} 
Subsequent mergers build what is likely to be the BCGs. 
At each step in the evolutionary chain, the black hole and 
spheroid may grow, with 
an entropy increase that sees pre-existing disc stars on ordered circular orbits 
become a part of a `dynamically hot' bulge, aka spheroid.  New stars can also
form, and the black holes can grow through black hole mergers, consumption of stars,
and gas-fuelling.

Previously, with just one recognised type of lenticular
galaxy\footnote{\citep{1925MNRAS..85.1014R} introduced the spiralless disc
galaxy type as a bridge between E and S galaxy types. As with the S
galaxies, the low-mass S0 galaxies need not contain a bulge component 
\citep[e.g.,][]{1976ApJ...206..883V}, although all in our sample do.}, 
prior to the realisation that the S galaxies 
are the bridging population between dust-poor and dust-rich S0 galaxies ---
the above sequence of galactic transformation had not been fully appreciated.
\citet{Graham-triangal} used the above sequence to modify the galaxy anatomy
sequence of Jeans-Reynolds-Lundmark-Hubble \citep{1919pcsd.book.....J,
  1920MNRAS..80..746R, 1925MNRAS..85..865L, 1925MNRAS..85.1014R,
  1926ApJ....64..321H, 1927UGC..........1L, 1936rene.book.....H} and also the
\citet{1976ApJ...206..883V} Trident and ATLAS$^{\rm 3D}$ Comb
\citep{2011MNRAS.416.1680C}.  Moreover, \citet{Graham-triangal} included
evolutionary pathways to produce a quasi-triangular-shaped sequence, referred
to as the `Triangal', that tracks growth in stellar mass, particularly the
spheroids' stellar mass.\footnote{\citet{Graham:Sahu:22b} discuss `non-growth'
  pathways which either halt star formation and turn S galaxies into gas-poor
  S0 galaxies \citep[e.g.,][]{1951ApJ...113..413S, 1972ApJ...176....1G} or
  strip stars to produce smaller galaxies \citep[e.g.,][]{1996Natur.379..613M,
    2001ApJ...557L..39B}.}  Here, we report on the presence of this growth
sequence in the diagram of specific star formation rate (sSFR) versus stellar
mass.

This present work follows 
\citet{2016ApJ...830L..12T} and \citet{2017ApJ...844..170T}, which interpreted the
red and blue sequence from \citet{2016ApJ...817...21S} as primarily a consequence of 
black hole feedback rather than galaxy morphology and, thus, assembly history.
Rather than showing the galaxy morphology in the $M_{\rm bh}$-$M_{\rm *,gal}$
diagram, those works presented the sSFR in this diagram.  They argued
that systems with lower $M_{\rm bh}/M_{\rm *,gal}$ ratios are less efficient
at shutting down star formation; that is, the AGN feedback was said to be less
efficient in galaxies with lower $M_{\rm bh}/M_{\rm *,gal}$ ratios, and
this explains why such galaxies have a higher sSFR.  At face value, this seems
plausible, and we investigate if this is the optimal interpretation.
We conclude it is not.

We make use of {\it Wide-field Infrared Survey Explorer} 
\citep[{\it WISE}:][]{2010AJ....140.1868W} data to measure or constrain the
SFRs of $\sim$100 galaxies with carefully considered (not automated) multicomponent
decompositions\footnote{These are superior to two-component
  \citet{1963BAAA....6...41S} bulge plus exponential disc
  \citep{1940BHarO.914....9P} decompositions.},
refined morphologies\footnote{Many S0 galaxies were previously mislabelled as E
  galaxies.}, and directly measured central black hole masses
(Section~\ref{Sec_data}). 
Given the proximity of the galaxy sample, typically within tens of Mpc, {\it
  WISE} can measure intrinsically low levels of dust emission and star
formation often missed in surveys. As such, our sample includes some low-mass
galaxies with low levels of star formation that are routinely undetected and
overlooked by large redshift surveys attempting to capture a representative
slice of the Universe. Similarly, morphology ``at a distance'' can be
challenging, so it too tends not to enter the star-formation conversation at a
level beyond LTGs versus ETGs. However, these facets prove to be essential for
understanding galaxy evolution as a whole and reflect a general need to
consider what the nearby Universe can uniquely tell us.

Section~\ref{Sec_Anal} explores the sSFR as a function of the 
$M_{\rm bh}/M_{\rm *,gal}$ ratio, the 
$M_{\rm bh}/M_{\rm *,sph}$ ratio, and the galaxy morphology.
While a broad trend between sSFR and $M_{\rm bh}/M_{\rm *,gal}$ is apparent,
the sSFR much more closely tracks the galaxy morphology. 
Similarly, while a trend between sSFR and $M_{\rm *,sph}/M_{\rm *,gal}$, i.e.,
spheroid-to-total stellar mass ratio, is also evident --- as seen in 
\citet{2009ApJ...707..250M} --- 
the primary connection again appears to be between the
sSFR and the galaxy morphology. 
Various sSFR-$M_*$ diagrams are presented in Section~\ref{Part2}, and 
the appearance of the `Triangal' \citep{Graham-triangal} is 
revealed in the sSFR-$M_{\rm *,gal}$ diagram.
A summary is provided in Section~\ref{Summ}.

\section{Data}
\label{Sec_data}

\subsection{The galaxy/SMBH sample}

The galaxy sample consists of 103 galaxies with directly measured SMBH masses
\citep[tabulated in][]{Graham:Sahu:22a} and multicomponent decompositions of
Spitzer Space Telescope \citep[{\it SST}:][]{2004ApJS..154....1W,
  2004ApJS..154...10F} 3.6~$\mu$m images.  Rather than automated fits of 2, 3,
or 4 \citet{1963BAAA....6...41S} $R^{1/n}$ functions, the decompositions were
based on physical components such as bars (fit with a Ferrer function), rings
(Gaussian function), bulges (S\'ersic function), inner and outer discs
(truncated, anti-truncated, and unbroken exponential function), and more. 
The stellar masses of the spheroids used in the figures pertain to either the bulge component,
when present, of the disc galaxies or (practically) the bulk of an elliptical
galaxy modulo small nuclear discs when present. That is, bars and inner discs
do not enter into the spheroid mass. 
References to updated galaxy distances --- used for the SMBH masses and
absolute magnitudes --- are provided in \citet[][their Table~1]{Graham:Sahu:22a}, as are
references for where the decomposition of each galaxy can be seen. These
references also 
successively note from where the original SMBH masses were taken.  Galaxy
model magnitudes were built from the sum of the galaxy component magnitudes
and agreed (<0.02 mag) with direct integration of the light profile.

Among the disc galaxies, i.e., the S and S0 galaxies, their spheroids display
a continuous trend between S\'ersic index $n$, a measure of the radial
concentration of their stars \citep{2001MNRAS.326..869T, 2005PASA...22..118G},
and the stellar luminosity/mass 
\citep{2020ApJ...903...97S, 
  2023MNRAS.519.4651H}.  These spheroids are distinct from inner discs and
(peanut shell)-shaped structures \citep[e.g.,][]{1972MmRAS..77....1D,
  1986AJ.....91...65J, 2016MNRAS.459.1276C}.  In our data set,
inner discs were modelled as such and (peanut
shell)-shaped structures were either modelled
\citep[e.g.,][]{2019ApJ...873...85D, 2019ApJ...876..155S} as separate
components, e.g., lenses \citep{2006AJ....132.1859B} or effectively folded
back into the bar component from which they emerged
\citep{1975IAUS...69..349H, 1981A&A....96..164C, 2005MNRAS.358.1477A,
  2018ApJ...852..133S}.  This modelling was performed using an isophotal analysis
technique \citep{1978MNRAS.182..797C} that measured the
radially-varying Fourier harmonic terms that describe the isophote's deviations from
ellipses due to, for example, buckled bars.
The corrected version \citep{2015ApJ...810..120C} 
of the {\sc Ellipse} task \citep{1987MNRAS.226..747J}
in the Image Reduction and Analysis Facility (IRAF) was used
to extract the (desired) symmetrical component of 
the two-dimensional galaxy image.  This has the added benefit of leaving
the non-symetrical disturbances in the residual image for further analysis,
such as, for example, the discovery of the shredded galaxy Nikhuli
\citep{2021ApJ...923..146G}, ,
and preventing such features from biasing the image decomposition into
regular/standard components. 
The spheroids were modelled with the \citet{1963BAAA....6...41S} $R^{1/n}$
function described in\citep{2005PASA...22..118G}. 
Further details of the spheroid extraction process are provided in 
\citet{2016ApJS..222...10S}, \citet{2019ApJ...873...85D}, 
 \citet{2019ApJ...876..155S}, and \citet{Graham:Sahu:22b}.

Stemming from the decompositions of the galaxy images --- which were supported by recourse to
kinematic profiles and maps, often revealing the presence of rotating discs in ETGs --- 
came knowledge of the galaxy morphologies: spiral, lenticular, and
elliptical. 
From cursory (qualitative) inspection of images, discs in ETGs are notoriously hard
to spot and, as discussed in \citet[][and references
  therein]{2019MNRAS.487.4995G}, have often been missed.
For example,
Fornax~A was historically classified as an E galaxy, but 
\citet{1965Natur.207.1282S} and 
\citet{1991rc3..book.....D} designated it a peculiar S0 galaxy with a disc and 
\citet{1995A&A...296..319D} presented the rotation of the galaxy's stars. 
Therefore, 
established quantitative measures of the images were used to detect the
presence and quantify the extent of the galaxies' discs. 
While a qualitative judgement on the presence of a
spiral was initially performed, a quantitative analysis of most of the spirals can be
found in \citet{2017MNRAS.471.2187D}.
Due to the proximity of the galaxies, there was little ambiguity at to whether
a galaxy was a spiral galaxy or not.  Although weak spiral patterns can exist
\citep[e.g.,][]{2000A&A...358..845J, 2003AJ....126.1787G}, there is 
nothing controversial with our designation of the spiral galaxies, in that it 
agrees with decades of literature.  Furthermore, no weak spirals 
were discovered in the residual (galaxy minus model) images.  The only point worthy of note is that
here, as in \citep{Graham-S0}, NGC~2974 (aka NGC~2652) and NGC~4594 (the Sombrero galaxy) are regarded
as S0 rather than S galaxies. 

To be clear, 
galaxies identified as having large-scale discs, i.e., discs that dominate the
light at large radii but have no spiral pattern, are denoted as S0
galaxies.
The S0 galaxies are invariably `fast rotators,' as revealed through kinematic
profiles and maps available in the literature. No recourse to star formation
rates was used to identify any galaxy's morphological type or the
`dust bins' (see later) for the S0 galaxies.  No recourse to clumpiness,
small-scale or otherwise, was used to establish the galaxies' morphological
types. While some S0 galaxies have a smooth appearance --- which has, in part,
led to some past misidentifications as E galaxies --- other S0 galaxies appear
relatively clumpy at optical wavelengths (e.g., NGC~524, Fornax~A,
Centaurus~A). As such,
(wavelength-dependent) clumpiness should be interpreted with care in the sSFR-$M_{\rm *}$ diagram as it
may not have the same finesse as the morphological type.  
The ES galaxies \citep{1966ApJ...146...28L,2019MNRAS.487.4995G}, 
found to have spiralless intermediate-scale discs, i.e., discs  
that do not dominate the light at large radii, are tabulated 
in \citet{Graham:Sahu:22b}, as were the cD and 
BCGs.  The elliptical galaxies, of course, have neither a large-scale nor
an intermediate-scale disc.

The dust-poor versus dust-rich S0 galaxies were readily identified from a
visual inspection of optical images, as reported 
in \citet{Graham-S0}, with most of the dust-rich S0 galaxies individually recognised in the literature 
as having been built from 
major gas-rich mergers involving S galaxies.  These morphological types are
used in the present analysis to help understand the distribution of sSFRs in
the $M_{\rm bh}$-$M_{\rm *,sph}$ diagram and the sSFR-$M_{\rm *,sph}$ and
sSFR-$M_{\rm *,gal}$ diagrams. Conceivably, the build-up (in the disc
galaxies) and then the decline (in 
the E galaxies) of the specific dust mass could also be used.  We do not, 
however, have this measure.  

Investigating the lenticular galaxies, \citet{Graham-S0} established four
`dust bins' based upon a personal inspection of Hubble Space Telescope images. 
The bins were denoted with the following codes:
\begin{itemize}
\item N for no visible signs of dust;
\item n for not much other than a nuclear dust ring/disc; 
\item y for a weak/mild level of widespread dust; and 
\item Y for lots of clearly visible widespread dust, i.e. dust-rich. 
\end{itemize}  
Examples can be seen in figure~1 of \citet{Graham-S0}. 
Once galaxies become massive, as in the case of the E and ES,e galaxies, they
are known to establish hot X-ray emitting halos of gas  which effectively destroy the dust
grains and eliminate the cold gas enclaves for dust 
\citep{1979ApJ...231..438D, 2003ApJ...599...38B, 2021A&A...649A..18G}. 
Among the sample of 37 E and ES,e galaxies, including the BCG matching this
description, only three are dust-rich (dust bin $=$ Y), while 19 have no visible dust, 8 have
only a nuclear dust disc, and 7 have only faint traces of dust beyond their 
nucleus. 

Two of the 103 galaxies (NGC~4395 and NGC 6926) are bulge-less and, as such, only
appear in diagrams showing the disc or galaxy stellar mass and
not in diagrams showing the bulges' stellar mass.\footnote{The terms bulge and
  spheroid are used interchangeably in this article.} 
Additional notes about the galaxies labelled in some figures are provided in
Appendix~\ref{Apdx1}.

\subsection{Stellar masses}
\label{subsec_mass}

\citet{Graham:Sahu:22a} tabulates the spheroid and galaxy stellar masses,
along 
with the SMBH masses, for the sample.  The stellar masses were obtained 
using the Spitzer 3.6 $\mu$m galaxy model magnitudes and 
colour-dependent (stellar mass)-to-light ratios based on realistic dusty
models created by \citet{2013MNRAS.430.2715I} 
for ``samples that include a range of morphologies, intrinsic colours
and random inclinations''.  The $M_*/L_{3.6}$ ratios\footnote{The
$M_*/L_{2.2}$ expression in \citet{2013MNRAS.430.2715I} was converted to
$M_*/L_{3.6}$ in equation~2 in \citet{Graham:Sahu:22a}.} 
are around 0.45 for the
LTGs and 0.7--0.9 for the ETGs \citep[][their figure~1]{Graham:Sahu:22a}, and 
the stellar masses were adjusted to 
the \citet{2002Sci...295...82K} initial mass function (IMF). 

\citet{2023ApJ...946...95J} present an alternate prescription to derive
stellar
masses based on W1 (3.4~$\mu$m) 'total flux' photometry and colours from {\it
  WISE}.
This approach was applied to the galaxy sample, many of which 
 were already included in \citet{2019ApJS..245...25J}.  These stellar masses 
were derived using a \citet{2003PASP..115..763C} 
IMF.  Subtracting 0.05 dex from these {\it WISE}-based 
masses effectively converts them to the \citet{2002Sci...295...82K}
 IMF \citep[][their table~2]{2010MNRAS.404.2087B}. 
 
\begin{figure}
\begin{center}
\includegraphics[trim=0.0cm 0cm 0.0cm 0cm, width=1.0\columnwidth,
  angle=0]{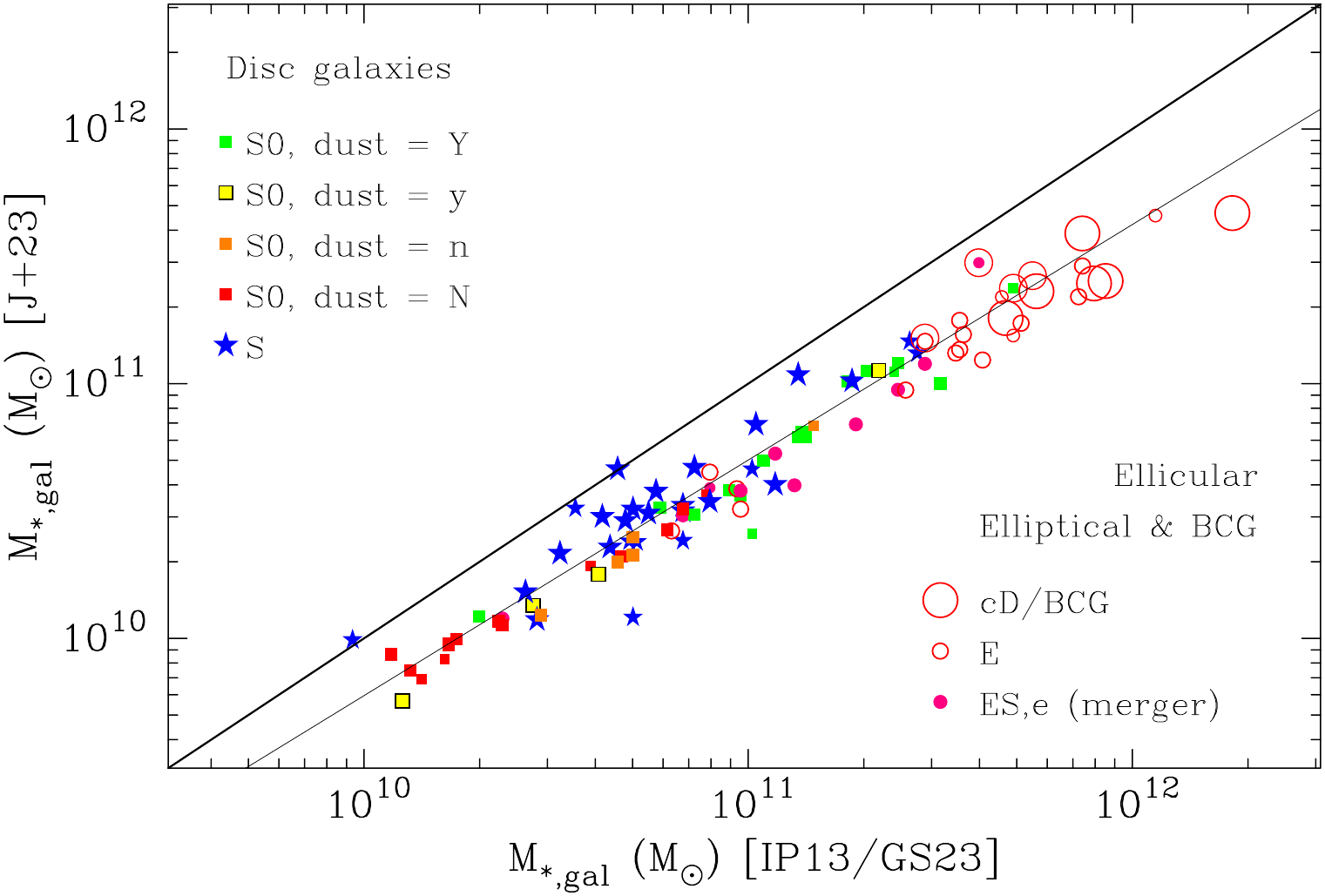}
\caption{Comparison of the galaxy stellar masses from (i) multicomponent fits 
  to {\it SST} 3.6~$\mu$m imaging data  \citep[][their
    table~1]{Graham:Sahu:22a} and using a slight 
modification of a stellar
  population model from \citet[][]{2013MNRAS.430.2715I}, and (ii) a
  derivation based on {\it WISE} photometry
  \citep[][]{2023ApJ...946...95J}
  after adjusting from a \citet{2003PASP..115..763C} IMF to a 
   \citet{2002Sci...295...82K} IMF (see Table~\ref{Table-data}). 
The line through the data is given by Eq.~\ref{Eq1}. 
}
\label{Fig1}
\end{center}
\end{figure}

Fig.~\ref{Fig1} compares the {\it WISE}- and {\it SST}-based 
stellar masses mentioned above.  They are seen to differ 
 by roughly a factor of two.  The offset has been approximated by the relation
\begin{equation}
\log(M_{\rm SST}/M_\odot) = (6/5.55)[\log(M_{\rm WISE}/M_\odot)-7]+7.
\label{Eq1}
\end{equation}
Such offsets have been seen before, e.g., by 
\citet[][their figure~7]{2015MNRAS.452.3209R}, 
\citet[][their figure~4]{2019MNRAS.484..814G}, 
and \citet[][their figure~4]{2019ApJ...876..155S}. 
They can stem from many issues, including 
the adopted stellar population model and, at infrared wavelengths, the
correction for dust glow. Some differences in stellar mass can have their
origin in the use of aperture photometry versus extrapolated model photometry from, for
example, a single S\'ersic $R^{1/n}$ fit or a two-component S\'ersic
$R^{1/n}$-bulge plus exponential-disc fit.
As \citet{2023MNRAS.518.1352S} revealed, 
it is crucial to be aware of such offsets when applying the 
$M_{\rm bh}$-$M_{\rm *,sph}$ and $M_{\rm bh}$-$M_{\rm *,gal}$ relations to
galaxy samples whose stellar masses may have been inconsistently measured. 
Differences in the derivation of the stellar mass between samples led to the
incorrect suspicion \citep{2016MNRAS.460.3119S} 
that ETGs with directly measured black hole masses are
offset from the ETG population at large in the $M_{\rm *,gal}$-(stellar
velocity dispersion, $\sigma$) diagram. 
For the current analysis, we do not need to 
track down the cause(s) of the difference observed in Fig.~\ref{Fig1}.
Nonetheless, it is noted that it primarily arises due to the adopted $M/L$
ratio. Unless otherwise specified, 
we proceed using the stellar masses from \citet{Graham:Sahu:22a}, upon which
the black hole scaling relations were previously derived. 

\begin{figure*}
\begin{center}
$
\begin{array}{ccc}
\includegraphics[trim=0.0cm 0cm 0.0cm 0cm, width=0.3\textwidth,
  angle=0]{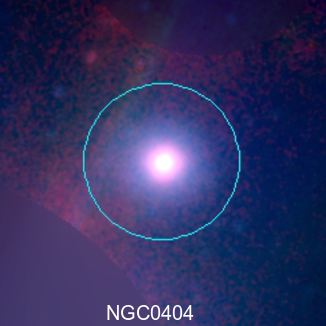} &
\includegraphics[trim=0.0cm 0cm 0.0cm 0cm, width=0.3\textwidth,
  angle=0]{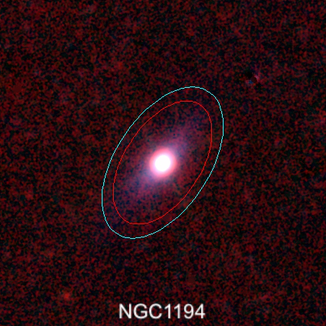} & 
\includegraphics[trim=0.0cm 0cm 0.0cm 0cm, width=0.3\textwidth,
  angle=0]{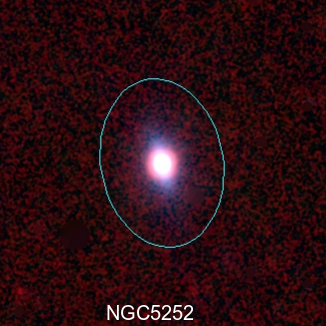} \\
\includegraphics[trim=0.0cm 0cm 0.0cm 0cm, width=0.3\textwidth,
  angle=0]{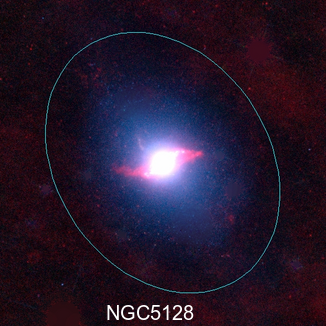} &
\includegraphics[trim=0.0cm 0cm 0.0cm 0cm, width=0.3\textwidth,
  angle=0]{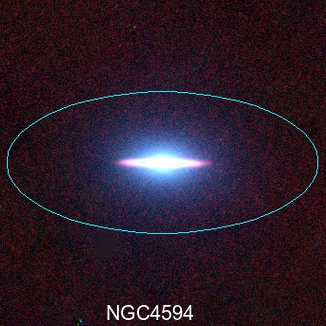} &
\includegraphics[trim=0.0cm 0cm 0.0cm 0cm, width=0.3\textwidth,
  angle=0]{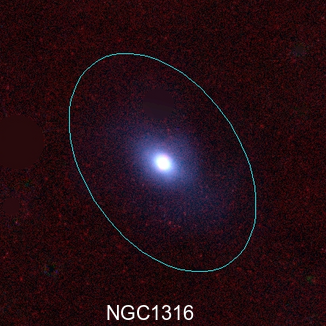} \\
\end{array}
$
\end{center}
\caption{
{\it WISE} 3-colour (W1, W2 and W3 bands) cutouts of select galaxies, 
after the photometry pipeline has
removed foreground stars, background galaxies and other corrections to achieve
a cleaner field for analysis.  
The ellipse denotes the photometric aperture to
extract the isophotal flux, roughly 23 mag arcsec$^{-2}$ 
(Vega) in W1 surface brightness.  This isophote represents the 1-sigma extent of the
galaxy. Total galaxy luminosities were obtained by going beyond this aperture,
using a double-S\'ersic model.
}
\label{Fig2}
\end{figure*}

\subsection{Star Formation Rates}
\label{Sec_SFR}

\begin{figure}
\begin{center}
\includegraphics[trim=0.0cm 0cm 0.0cm 0cm, width=1.0\columnwidth, angle=0]{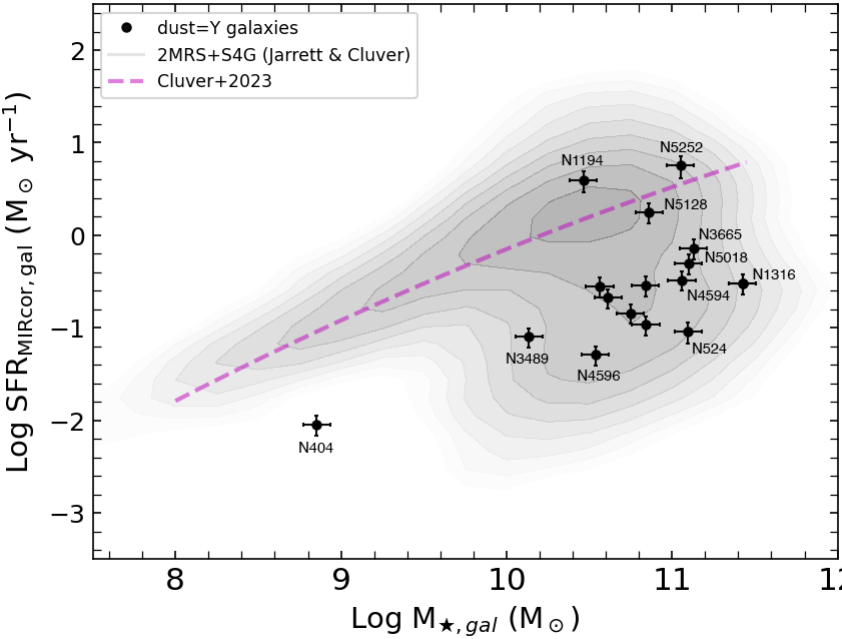}
\caption{Star formation rate based on {\it WISE} mid-infrared measurements
  calibrated to the total infrared luminosity (Cluver$+$2023 = Cluver et
  al.\ in preparation) versus the {\it WISE}-based stellar mass.  The SFRs,
  given 
  in Table~\ref{Table-data}, are shown here for just the dust-rich E/S0 galaxies in the
  dust bin `dust$=$Y'. They do not simply cluster about 
  the high density centre of the ~red sequence/cloud. The relative location of
  the dust-poor S0 and other galaxy types will be seen in later figures.
  To cover the region occupied by normal galaxies
  (Jarrett et al.\ in preparation), the contours pertain to a larger
  redshift-zero sample that includes the 2MASS Redshift Survey (2MRS) and
  Spitzer Survey of Stellar Structure in Galaxies (S$^4$G).
The magenta line represents the 
`main sequence' as derived by Cluver et al.\ (in preparation)
using the combined S$^4$G and 2MRS samples.
}
\label{Fig3}
\end{center}
\end{figure}

The level of star formation activity in the past 100~Myr is assessed 
through the infrared
emission as measured using the {\it WISE} total integrated fluxes and the
calibration developed by \citet[][2023 in preparation]{2017ApJ...850...68C}, 
which shows that the 12 and 23~$\mu$m
mid-infrared bands of {\it WISE} are an effective predictor
of the total infrared luminosity.  Half a dozen example galaxy images from
{\it WISE} are shown in Fig.~\ref{Fig2}. The selection of these six will
make sense upon reading Appendix~\ref{Apdx2}. 
The mid-IR is a robust tracer of the dust-obscured star formation activity,
but less so with lower mass galaxies that tend to be more dust-free and
metal-poor.  Cluver et al.\ (in preparation) have created a UV-based correction
to this regime, generally applicable to dwarf galaxies, $\log(M_*/M_\odot) <
10$ dex, which accounts for the lower 
dust opacity and hence, can be used to estimate the total SFR. 
However, this is only relevant to one galaxy in the sample. 

The SFR ($M_\odot$ year$^{-1}$) is derived from the mid-infrared
spectral ($\nu$L$_\nu$) luminosity, 
as measured in the {\it WISE} W3 (12~$\mu$m) and W4 (23~$\mu$m) imaging bands
after
subtracting the stellar continuum \citep[inferred from the W1 integrated
  flux:][]{2017ApJ...850...68C, 2019ApJS..245...25J}.
Cluver et al.\ (in preparation) 
provide the scaling relations between the mid-IR luminosities and the total
infrared luminosity (LTIR),
updated from \citet{2017ApJ...850...68C}, which is then converted to the SFR using
\citet{2013seg..book..419C}.  The mid-IR SFR
is then a combination of the inverse-variance weighted W3 and W4 SFRs to
derive the optimal SFR traced by the WISE mid-IR bands. 
In this paper, the SFRs for the sample
galaxies are compared to their host stellar mass using the galaxy
`main-sequence' diagram: global SFR vs mass.  A useful variation of this
diagram uses the sSFR to gauge 
the rate at which galaxies are building their disc/bulge 
(relative to past star formation), for example, \citet{2018ApJ...859...11S}. 

In Fig.~\ref{Fig3}, 
the SFR for the dust-rich S0 galaxies is shown against the {\it WISE}-based galaxy stellar mass 
\citep{2023ApJ...946...95J}. 
The bulk of the dust-rich S0 galaxies can be seen to
occupy the region for ETGs below the 
`main sequence' (magenta curve) that normal, star-forming
galaxies tend to occupy given their global (past-to-present) stellar mass.
However, they do not simply cluster about the high-density region of the red
cloud/sequence centered at $\log(M_{\rm *,gal}/M_\odot) \sim 11$ dex and $\log({\rm
  SFR}\, M_\odot\, {\rm yr}^{-1})\sim -1$ dex, where the E/ES,e galaxies will later be
shown to reside.

Not all of the galaxies in the sample are detected in {\it WISE}'s W3 or W4
bands, notably after the stellar continuum contribution has been subtracted
from the integrated fluxes. From the 103 galaxies, we were unable to measure
an SFR, or rather, we only have upper-limit measurements for one ES,e galaxy,
five dust-poor (dust$=$N) S0 galaxies, and 14 E galaxies.  While this is good,
we can turn this statement around to report that 9 of 10 ES,e and 9 of 14
dust-poor (dust$=$N) S0 galaxies, plus 11 of 25 E galaxies, show up with a
non-zero SFR.  As revealed in the figures, the associated sSFR rates are at
the low level of $10^{-13} \lesssim {\rm M}_\odot {\rm yr}^{-1} \lesssim
10^{-12}$.

In passing, it is noted, albeit with a sample of just four ES,b galaxies
(treated herein as S0 galaxies), that the ES,b galaxies overlap with the ES,e
ellicular galaxies in the sSFR vs $M_{\rm *,galaxy}$ diagram.

It is also observed that the E galaxy IC~4296 has an undigested component,
i.e., a captured galaxy, containing 
4.5 per cent of the total stellar mass \citep{2019ApJ...876..155S}. The past merger
giving rise to this component may have induced some star formation, as IC~4296
has a sSFR of $-$12.74 dex.
The E galaxy NGC~4261 may also contain an undigested component, 
perhaps explaining why it has an sSFR of $-$12.36 dex. 
However, there are nine other E galaxies with a non-zero SFR for which our galaxy decomposition
did not require the inclusion of such a component. It could be that such
yet-to-be-assimilated components are (now) small
enough not to require modelling.  Alternatively, 
the low sSFRs may be suggesting that something more is afoot, 
influencing the measurement.  These
galaxies may have some very low levels of warm dust emission heated by old
stars --- with the dust coming from
stellar winds \citep[e.g.,][]{1988A&A...206..153G, 2004ApJ...608..405C} and
supernovae \citep[e.g.,][]{2001MNRAS.325..726T,2003ApJ...592..404L,
  2011Sci...333.1258M} --- as opposed to heating from recent ($\sim$100 Myr) 
star-formation. 
This emission is typically negligible relative to young-disc star formation emission but
may be appreciable for the stellar-dominated (i.e., bulge) components 
of ETGs \citep{2019ApJS..245...25J}.  We therefore caution 
that some of the {\it WISE} SFRs are likely over-estimated and may serve best as
upper limits.

\subsubsection{Active Galactic Nuclei}

\begin{figure}
\begin{center}
\includegraphics[trim=0.0cm 0cm 0.0cm 0cm, width=1.0\columnwidth, angle=0]{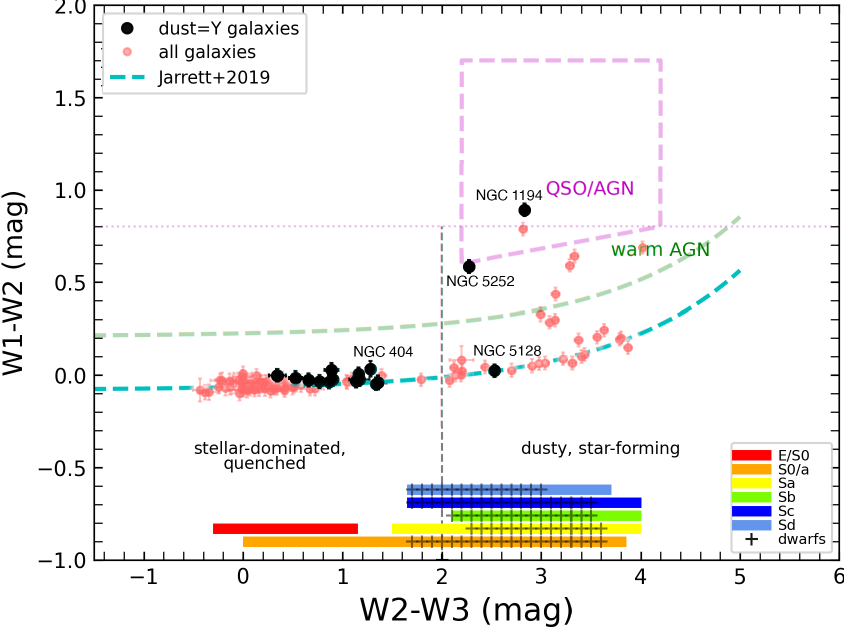}
\caption{{\it WISE} colour-colour plot for``all''
  galaxies (red points). The S0 galaxies with dust$=$Y are highlighted
(black points). All colours are tabulated in Appendix~\ref{Apdx3}. 
% Highlighted cases:   
NGC~1194 and NGC~5252 are Seyferts with warm W1-W2 colours. Although NGC~5128
(Cen~A) is particularly dusty in the central region, its total
host galaxy colour is nominal (see also Fig.~\ref{Fig2}).
The diagram indicates where normal star-forming galaxies 
are expected to lie \citep[dashed thick cyan curve from
  Jarrett$+$2019 = ][]{2019ApJS..245...25J}, with most dusty ETGs 
on the left and active star-forming galaxies on the right.
Five spiral galaxies (Circinus, NGC~1320, NGC~4151, NGC~4388, and NGC~7582) appear
above the dashed thin green curve for AGN, with three galaxies 
(NGC~1275, NGC~2273, and NGC~3227) just below it.
Vertical deviations indicate ``warm'' colours, typically associated with
AGN activity, at low levels (see green dashed curve) and where the 
AGN dominates the total emission (magenta dashed box and dotted line).
}
\label{Fig4}
\end{center}
\end{figure}

Star formation activity is inferred from the continuum-subtracted {\it WISE}
luminosity.  The underlying assumption is that the infrared emission arises
from interstellar medium dust/molecules heated by UV radiation from recent
star formation.  Complications arise from dust opacity and metallicity
(as noted above), dust heating from old stars (unrelated to recent star formation), and
active galactic nuclei, whose dusty torus emission may be powerful enough
to compete with or even overwhelm the host galaxy emission.  
As measured in mid-IR, the galaxy colours 
may be used to detect dust-obscured AGN emission
contaminating the star formation tracers \citep{2011ApJ...735..112J, 2019ApJS..245...25J,
  2022ApJ...939...26Y}.

A couple of galaxies should be, and are, disregarded due to dust-heating from
their AGN: 
they are NGC~1275 
and the Circinus 
galaxy.  They have artificially high sSFRs, evident in later figures.  
Among the 17 dust-rich S0 galaxies (dust=Y),  
a few appear to have elevated SFRs, placing them on the 
star-forming main sequence, as seen in Fig.~\ref{Fig3}.
None of the four dust$=$y S0 galaxies reside above this main sequence. 

Fig.~\ref{Fig4} shows the {\it WISE} W1$-$W2 colour versus the W2$-$W3 colour
for these dust-rich (dust=Y) S0 galaxies. 
A relatively red W1$-$W2 colour (top-side of the diagram) is a signature of
AGN heating dominating the global mid-infrared fluxes. It is 
problematic because it can elevate the inferred SFR.  This is, however, not
responsible for the bulk of the S0 galaxies with low-to-moderate SFRs.
They reside 
below the main sequence of star formation, delineated by the magenta curve in Fig.~\ref{Fig3}.
Some reside in the so-called `green valley', between this blue sequence of star-forming
galaxies and the red cloud/sequence occupied by passive galaxies.  
Apart from NGC~1194 and NGC~5252, 
both of which appear to harbour dust-obscured AGN, 
the dust-rich S0 galaxies do not have colours suggestive of a warm AGN
component 
or a QSO with (W1$-$W2)$>$0.4 mag.  None of the other S0 galaxies contains a warm
AGN. 
This means that the SFRs are reliable and unlikely a result of AGN-heating
of the dust \citep[e.g.,][]{2000AJ....119..991S, 2015ApJ...814....9K}.

It is noted that in the past, the entire star-forming main sequence was more active, with 20 
times more star formation at a redshift of 1 and 40 times more at $z=4$
\citep{2015A&A...581A..54T}, see also 
\citep{2004MNRAS.355..374B, 2010MNRAS.407..830M, 2023arXiv230408516C, 
2023MNRAS.519.1526P, 2023NatAs...7..611R}. 
It is unsurprising to find AGN activity in gas-rich mergers, which may
indicate how recently a disturbance or merger occurred.  Indeed, due to
angular-momentum-zapping gravitational tidal forces sending gas inward to fuel
the central black hole(s), the dusty S0 galaxies likely contained quasars 
\citep[e.g.,][]{2008ApJS..175..356H} and may have briefly became `starbursts' 
and ultraluminous IR galaxies (ULIRGs) with
sSFR $>10^{-9}$ M$_\odot$ yr$^{-1}$ when they first formed \citep{1992ApJ...400..153M}.
In the distant past, when today's dust-poor S0 galaxies were forming on the
star-forming sequence, a collision between two of them would also have led to
a starburst and perhaps a ULIRG if high-angular-momentum gas reservoirs were
brought into play.

In Appendix~\ref{Apdx2}, we continue the discussion about AGN with
information on individual galaxies.

\begin{figure*}
\begin{center}
\includegraphics[trim=0.0cm 0cm 0.0cm 0cm, width=1.0\textwidth, angle=0]{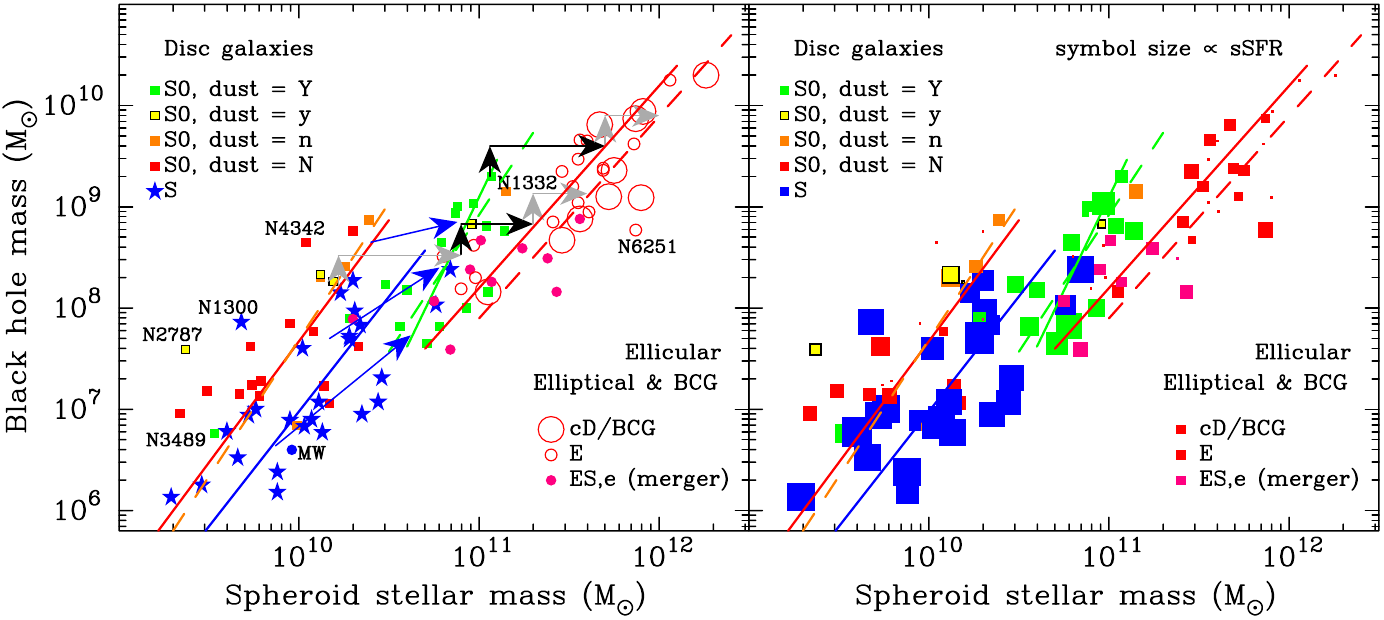}
\caption{
Left: $M_{\rm bh}$-$M_{\rm *,sph}$ diagram as a function of galaxy
  morphology \citep[figure taken from][]{Graham-triangal}. 
The red solid line pertains to E and ES,e galaxies, while the red dashed line is representative of the cD and BCGs.
  Right:  $M_{\rm bh}$-$M_{\rm *,sph}$ diagram as a function of galaxy
sSFR$_{\rm *,galaxy}$. Here, the symbol size reflects the logarithmic value of
sSFR$_{\rm *,galaxy}$. 
Only Circinus (S) and NGC~1275 (E) have been excluded from the figure,  
given how their AGN-heated dust inflates their measured sSFR --- see
Fig.~\ref{Fig8}. 
As explained in \citet{Graham-triangal}, the labelled galaxies are shown 
but excluded from the Bayesian regression analyses performed there.
Theories for the emergence of spiral patterns in galaxy discs, and evidence of
acretion and (bulge-building) minor mergers in S galaxies, is noted in
\citet{Graham-triangal}.
The dust-rich S0 galaxies are recognised major wet merger products
while the E galaxies, typicaly with depleted cores, are major dry merger
remnants \citep{Graham-S0}. At a fixed $M_{\rm bh}$, both $M_{\rm
  *,sph}$ and $M_{\rm *,gal}$ \citep[][appendix fig.~A4]{Graham-triangal} in
dust-poor S0 galaxies are lower than that in S galaxies, disfavouring the
faded/transformed S origin for most dust-poor S0 galaxies.
}
\label{Fig5}
\end{center}
\end{figure*}

\begin{figure}
\begin{center}
\includegraphics[trim=0.0cm 0cm 0.0cm 0cm, width=1.0\columnwidth, angle=0]{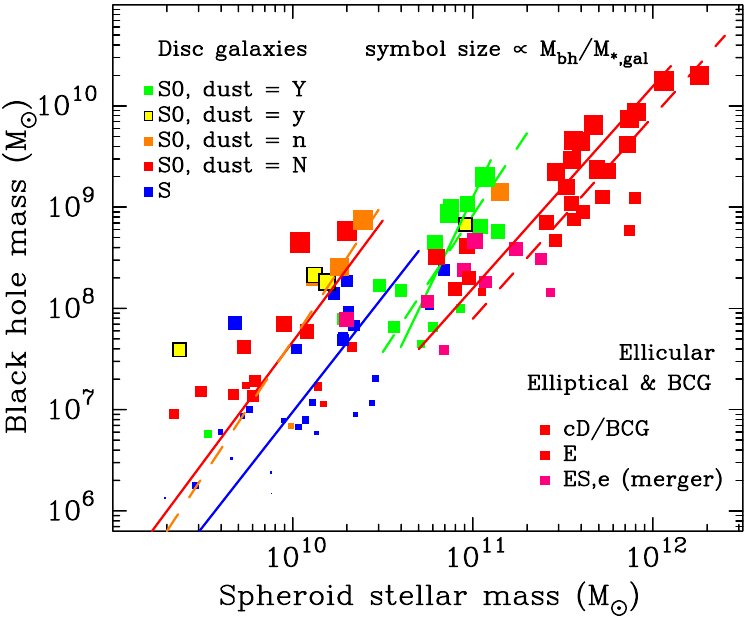}
\caption{
Similar to the right-hand panel of Fig.~\ref{Fig5}, but 
here the symbol size represents the $M_{\rm bh}/M_{\rm *,gal}$ mass ratio,
previously claimed to be regulating the sSFR \citep{2016ApJ...830L..12T}.
This is tested in Fig.~\ref{Fig7} against the notion that the
galaxy morphology, reflective of the galaxy accretion/merger history, is
the driving force.
}
\label{Fig6}
\end{center}
\end{figure}

\section{Analysis and Discussion}

\subsection{Part 1: SFRs, AGN feedback, and galaxy morphology in the $M_{\rm bh}$-$M_{\rm *,sph}$ diagram}
\label{Sec_Anal}

\begin{figure*}
\begin{center}
\includegraphics[trim=0.0cm 0cm 0.0cm 0cm, width=1.0\textwidth, angle=0]{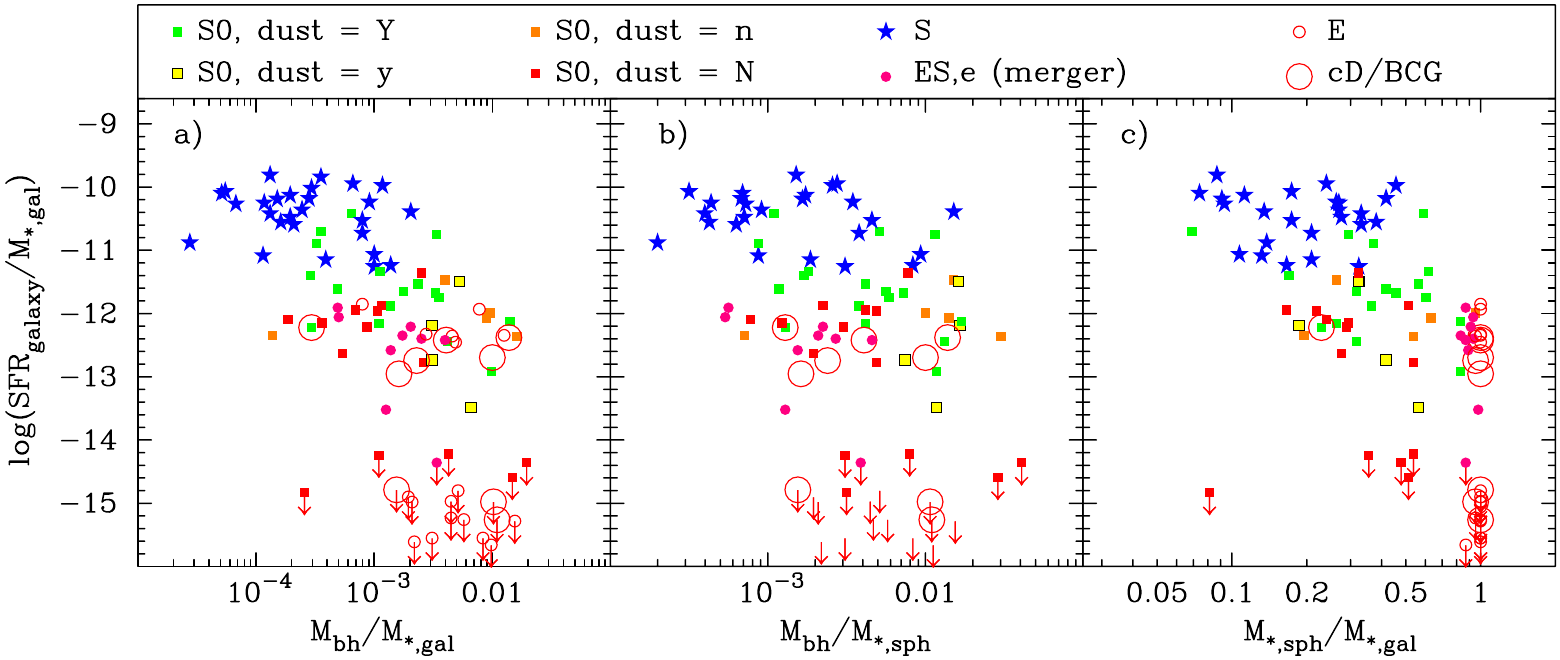}
\caption{
Galaxy specific-SFR versus 
a) specific black hole mass (left-hand panel), 
b) $M_{\rm bh}/M_{\rm *,sph}$ (middle panel), and 
c) $M_{\rm *,sph}/M_{\rm *,gal}$ (right-hand panel).
Note that the trend in the left-hand panel is equivalent to that in the SFR versus black
hole mass diagram.  It is apparent that the SFR correlates better with the
galaxy morphology than the BH-to-stellar mass ratio or the bulge-to-total
stellar mass ratio (right-hand panel). 
Circinus and NGC~1275 have been excluded from the figure for clarity, given
that their AGN-heated
dust inflates the measured sSFR. Low-mass dust-poor S0 galaxies have low
central surface brightness and are likely 
missing from the lower left due to sample selection. 
}
\label{Fig7}
\end{center}
\end{figure*}

Fig.~\ref{Fig5} presents an enhanced view of the galaxy-morphology-dependent 
$M_{\rm bh}$-$M_{\rm *,sph}$ diagram, previously shown in terms of E, S0 and S
galaxies \citep{2016ApJ...817...21S, 2019ApJ...876..155S}.  The broad and
roughly quadratic trend seen at $M_{\rm *,sph} \lesssim 10^{11}$ M$_\odot$ for the
ensemble of disc galaxies (S galaxies and dust-rich and dust-poor S0 galaxies) 
was highlighted a decade ago by separating off the E galaxies (built from dry
mergers and) possessing cores depleted of stars due to binary SMBHs 
\citep{2012ApJ...746..113G, 2013ApJ...764..151G,
  2013ApJ...768...76S}.  Following \citet{2016ApJ...830L..12T}, see also
\citet{2017ApJ...844..170T}, the right-hand panel of Fig.~\ref{Fig5} displays 
the sSFRs (Section~\ref{Sec_SFR}).  
It is apparent that the S galaxies have the highest sSFRs, followed
by the dust-rich S0 galaxies.  The dust-poor S0 and E galaxies (including ES,e
and BCG) have low to no detectable sSFR.

\citet{2016ApJ...830L..12T} suggested that the $M_{\rm bh}/M_{\rm *,gal}$ mass
ratio dictates the strength of the AGN feedback and, in turn, regulates the
sSFR through quenching. 
To help visualise this scenario, Fig.~\ref{Fig6} displays the $M_{\rm
  bh}/M_{\rm *,gal}$ mass ratio instead of the sSFR seen in Fig.~\ref{Fig5}.
Two patterns are evident.  As one moves down the (galaxy morphology)-dependent
sequences to lower masses, the $M_{\rm bh}/M_{\rm *,gal}$ ratio decreases.
This is as expected given the near-quadratic slope of these $M_{\rm
  bh}$-$M_{\rm *,sph}$ relations and the equal or steeper $M_{\rm bh}$-$M_{\rm
  *,gal}$ relations \citep{Graham-triangal}.  The second trend is the reduced
$M_{\rm bh}/M_{\rm *,gal}$ ratio at fixed BH mass as one moves to higher
$M_{\rm *,sph}$.  By definition, this is expected for the E galaxies in which
$M_{\rm *,sph} \approx M_{\rm *,gal}$.  For the disc galaxies, it is simply a
case of more massive spheroids with the same black hole mass being associated
with more massive galaxies.

Modulo a suspected sample selection bias which has missed low-mass BHs in
low-stellar-mass, low central surface brightness, dust-poor S0 galaxies,
Fig.~\ref{Fig7}a reveals a trend of decreasing sSFR with increasing `specific
BH mass', $M_{\rm bh}/M_{\rm *,gal}$. This trend was reported by
\citet{2017ApJ...844..170T} and is seen in Fig.~\ref{Fig6}.  
Multiplying both axes by $M_{\rm *,gal}$ in Fig.~\ref{Fig7}a gives 
an identical trend of decreasing SFR with increasing BH mass.
This behaviour is, therefore, understood because elliptical galaxies --- built from the merger
of disc-dominated galaxies --- have larger BH masses and lower star formation
rates than spiral galaxies. 
The future inclusion of low-stellar-mass ETGs with
$M_{\rm bh}\sim10^{6\pm1}$ is anticipated to alter this trend. 
Extrapolating the $M_{\rm bh}$-$M_{\rm *,gal}$ trends seen in
\citet[][figure~A4]{Graham-triangal}, these disc-dominated
low-mass ETGs with low SFR would extend the shelf of dust-poor galaxies
(coded orange and red) in Fig.~\ref{Fig7} to lower $M_{\rm bh}/M_{\rm *,gal}$
ratios, thereby reducing the strength of the correlations seen there.

\citet{2017ApJ...844..170T} interpreted the observed trend as indicating a
suppression of the sSFR due to greater AGN feedback in systems with higher
specific BH masses or, equivalently, a decreased SFR in systems with more
massive black holes due to increased AGN feedback.  However, and this is the
key, Fig.~\ref{Fig7} 
reveals that the sSFR$_{\rm *,galaxy}$ correlates better with the galaxy
morphology than it does with either of the BH-to-stellar mass ratios shown
there, as is evident from the scatter at fixed $M_{\rm bh}/M_{\rm *,gal}$ or
$M_{\rm bh}/M_{\rm *,sph}$.
That is, Fig.~\ref{Fig7} reveals that the sSFR cares more about the galaxy
morphology, which reflects the galaxy formation history, than the
$M_{\rm bh}/M_{\rm *,gal}$ (or $M_{\rm bh}/M_{\rm *,sph}$, Fig.~\ref{Fig7}b)
ratio.  In other words, ($M_{\rm bh}/M_{\rm *,gal}$)-dependent AGN feedback is
not the driving force regulating the SFRs in galaxies.  Accretions and
mergers, which set the morphology, are more critical.

A consequence of this result is that galaxy formation models should not tune
the sSFR primarily through prescriptions of black hole feedback tied to the
$M_{\rm bh}/M_{\rm *,gal}$ ratio but rather through (gas-rich/poor) accretion
events and mergers which dictate the galaxies morphologies
\citep{Graham-triangal}, coupled with, of course, gas recycling from stellar
winds \citep[e.g.,][]{2010ApJ...714L.275M}. 

Fig.~\ref{Fig7}c reveals that while the sSFR correlates more tightly with the
bulge-to-total, $B/T$, stellar mass ratio, $M_{\rm *,sph}/M_{\rm *,gal}$, than
the BH-to-stellar mass ratios (using either $M_{\rm *,sph}$ or $M_{\rm
  *,gal}$), the $B/T$ ratio is also not the driving force.\footnote{The same
  finding is reached when using the bulge-to-disc, $B/D =
  [(B/T)^{-1}-1]^{-1}$, stellar mass ratio, $M_{\rm *,sph}/M_{\rm *,disc}$.}
For example, dust-poor ETGs with $0.15<B/T<0.4$ have $-12 \lesssim sSFR_{\rm
  *,galaxy} \lesssim -13$ dex, while LTGs with the same $B/T$ ratios have $-10
\lesssim sSFR_{\rm *,galaxy} \lesssim -11$ dex.  This observation would appear
to be at odds with the idea that a bigger (relative to the disc) bulge
prevents star formation by stabilising the disc
\citep[e.g.,][]{2009ApJ...707..250M}.  While we do not rule out that this may
occur at some level, and it is again noted that inner discs, bars and
(peanut-shell)-shaped structures are excluded from our `bulge' mass, there
appears to be a bigger factor at play. Specifically, the galaxy morphology, a
tracer of the accretion and merger history, seems to dominate in
setting the sSFR.

\begin{figure*}
\begin{center}
\includegraphics[trim=0.0cm 0cm 0.0cm 0cm, width=1.0\textwidth, angle=0]{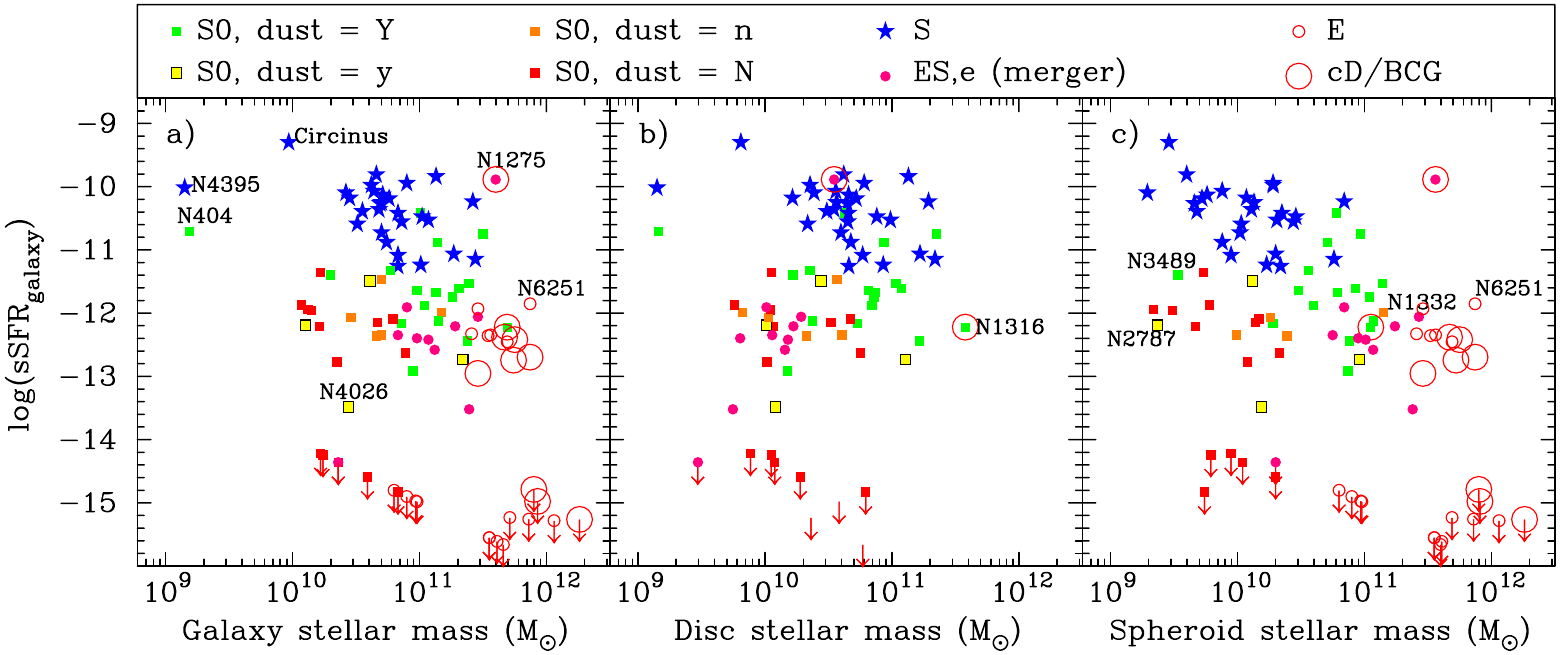}
\caption{Left: Galaxy sSFR$_{\rm galaxy}$ versus $M_{\rm *,galaxy}$.
The dust-poor S0 galaxies reside to the lower-left of the S-(dust-rich S0)-E
SFR main sequence. As indicated by the data points that are upper limits,
there is an artificial absence of systems in the lower-left of this diagram
due to limitations of detection of low sSFR in low-mass galaxies. 
Middle: Galaxy sSFR$_{\rm galaxy}$ versus $M_{\rm *,disc}$. 
While most BCGs in the sample are E galaxies, 
NGC~1316 is a dusty S0 galaxy, and NGC~1275 is an ES,e galaxy. 
Right: 
The bridging nature of the dust-rich S0 galaxies, between the S and E
galaxies, is more apparent here when using the spheroids' stellar mass on the
horizontal axis.
}
\label{Fig8}
\end{center}
\end{figure*}

\begin{figure}
\begin{center}
\includegraphics[trim=0.0cm 0cm 0.0cm 0cm, width=1.0\columnwidth, angle=0]{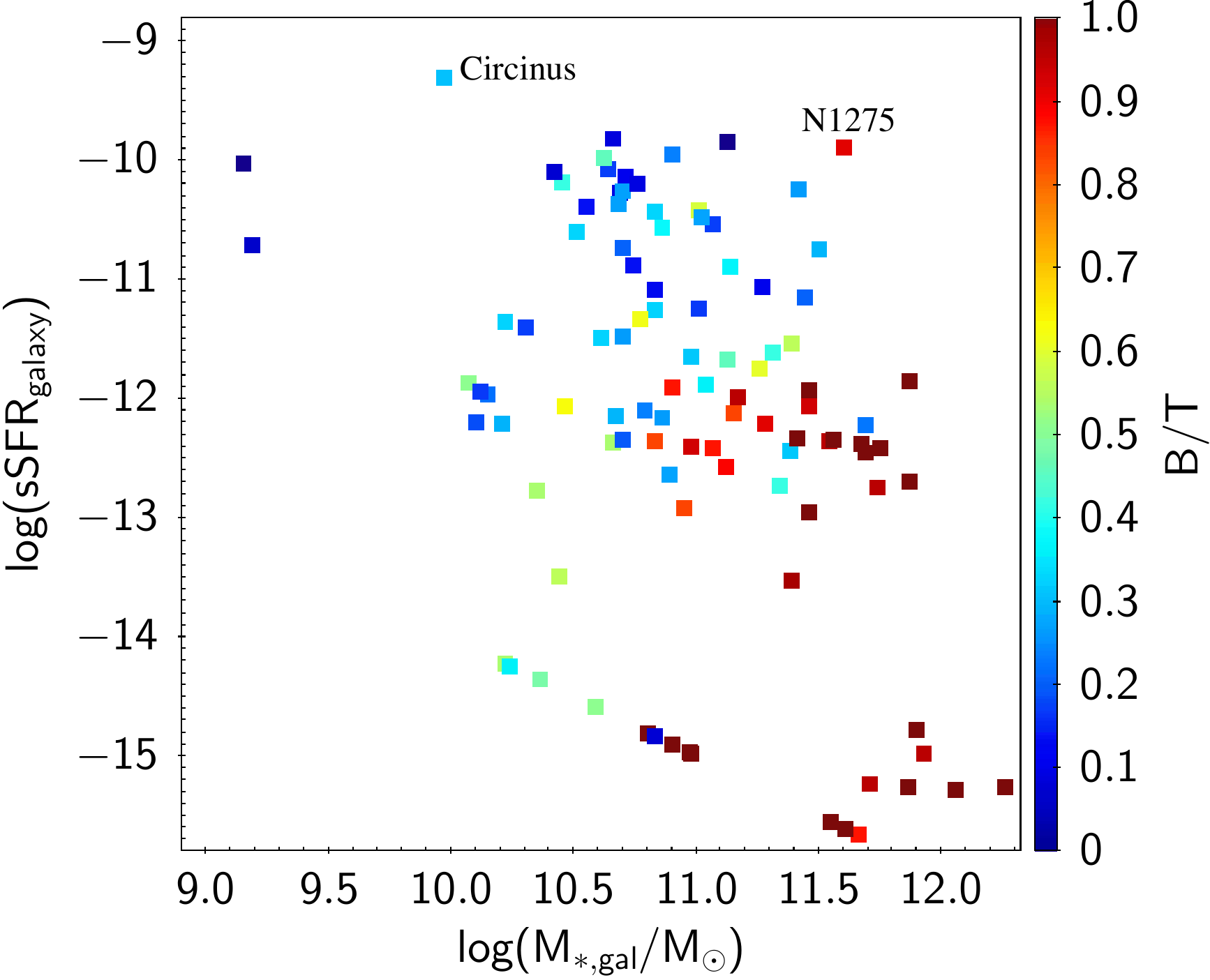}
\caption{Similar to the left panel of Fig.~\ref{Fig8} but showing the
bulge-to-total ($B/T$) flux ratio rather than the morphological type. As in
Fig.~\ref{Fig8}, sSFRs below $10^{-14}$ are upper limits.
In the upper right, the high $B/T$
galaxy NGC~1275 has an artificially high sSFR due to
dust-heating from its AGN, as does Circinus, which has the highest sSFR of the
sample. 
}
\label{Fig9}
\end{center}
\end{figure}

\subsection{Part 2: sSFR sequences}
\label{Part2}

In a sense, we are presenting two papers in one.  However, the connections are
sufficiently tight, and the data is the same, so the ideas are presented together.. 
In what follows, we extend the sSFR-(mass ratio) diagrams in Fig.~\ref{Fig7} to the
sSFR-(stellar mass) diagrams in Fig.~\ref{Fig8}.

Using far-IR {\it Herschel Space Observatory} \citep{2010A&A...518L...1P} data
and stellar mass estimates from the MAGPHYS SED-modelling program
\citep{2008MNRAS.388.1595D}, \citet{2018MNRAS.473.3507E} advocate for a single
Galaxy Sequence to replace/unite the `blue sequence' and `red cloud' in the sSFR-$M_{\rm
  *,gal}$ diagram.  Our data support this galaxy sequence --- of which the
high-mass end can be seen in Fig.~\ref{Fig8}a 
and, for example, \citet[][their figure~6]{2009MNRAS.393.1302W}, 
--- but also reveals
that this is not the complete picture.  The dust-poor S0 galaxies are missing from
this single Galaxy Sequence. They reside in a part of the diagram where sample
selection effects make them difficult to detect due to their lower stellar mass and low
dust content. Their absence is not a new result; \citet{2018MNRAS.473.3507E}
note that their survey missed these low-mass ETGs. 
However, because they are nearby, our {\it WISE} data detect many of them with $-12 \lesssim
sSFR_{\rm *,galaxy} \lesssim -13$ dex, while others only have upper limits
with $sSFR_{\rm *,galaxy} \lesssim -14$ dex (Fig.~\ref{Fig8}a). 
While \citet[][see their figure~10]{2017MNRAS.471.2687B} does not identify
primordial dust-poor S0 galaxies as separate from wet-merger-built dust-rich
S0 galaxies, they do observe the sSFR to track the E/S0/early-type S/late-type
S galaxy morphology.\footnote{A greater fraction of low-mass
($<3\times10^{10}$ M$_{\odot}$) S0 galaxies in the sample of
\citet{2017MNRAS.471.2687B} would have modified their results for quenched
galaxies, with S0 galaxies no longer predominantly occurring in the `green
valley', but their observation that the `green valley' is dominated by S0
galaxies and early-type S galaxies would still hold.}

The dust-rich ETGs/S0 galaxies, which are more massive, appear red in the
optical colour-magnitude diagram \citep{Graham:Sahu:22a}.  This is because the
dust obscures (and is heated by) the blue and UV light from the regions of
younger stars \citep[e.g.][]{2009ApJ...690.1883G}.  
With mid-IR observations focussing on the warm glow of the
dust, we have a better estimate of the SFR than optical (wide-band) colours
would provide.  This has allowed the so-called `green valley'
\citep[e.g.,][]{2007ApJ...665..265F, 2007ApJS..173..315S, 2007ApJ...658..161Z}, usually
a zone with low occupation numbers, to bloom \citep{2017ApJ...844...45O}.  The
mid-IR observations apparently uncover more of these transitional galaxies,
with implications for our understanding of valley crossing times.
The dust-rich S0 galaxies are likely the inhabitants of the `green mountain'
reported by \citet{2018MNRAS.473.3507E} and \citet{2018MNRAS.481.1183E} to
form a substantial population/bridge between the S and E galaxies, yielding
the single Galaxy Sequence that they report.  That is, we associate the
abundance of `green' galaxies primarily with major wet mergers rather than,
for example, fading S galaxies \citep{1951ApJ...113..413S,
  1972ApJ...176....1G, 1980ApJ...237..692L}, bulge
growth via secular evolution in S galaxies \citep{2017MNRAS.466.2570B}, or
AGN feedback \citep[e.g.][]{2016MNRAS.455L..82L, 2016ApJ...830L..12T}. 

In passing, we note that while the removal of gas and fading of S galaxies
will not reduce their spheroid's stellar mass, i.e., will not shift them to
the left in Fig.~\ref{Fig5}a, there are a few dust-poor S0 galaxies on the S
galaxy $M_{\rm bh}$-$M_{\rm *,sph}$ relation, near the Milky Way and with
intermediate sSFRs (Fig.~\ref{Fig5}). These could be gas-stripped, strangled
and failed S galaxies \citep[e.g.,][]{1951ApJ...113..413S,
  2006A&A...458..101A, 2008ApJ...672L.103K, 2017MNRAS.471.2687B,
  2022MNRAS.513..389R}, different to the bulk of the dust-poor S0 galaxies and
the quenching merger products, i.e., the dust-rich S0 galaxies.
It would also be of interest to know if the galaxies with centrally-located 
low ionisation emission line regions, found in the `green valley' 
\citep{2017MNRAS.466.2570B}, might be 
misidentified dust-rich S0 galaxies rather than S galaxies.

Fig.~\ref{Fig8}a reveals that, on its own, the galaxies' stellar mass is not
responsible for reducing the sSFR, a notion explored by
\citet{2010ApJ...721..193P}.  Conceivably, if this `mass quenching' operated
in tandem with `environmental quenching', it could match the pattern seen in
Fig.~\ref{Fig8}a if the dust-poor S0 galaxies reside in clusters.  While the
Appendix of \citet{Graham-S0} provides the environment of the S0 galaxies used
here, further detail, such as the presence/nature of a hot X-ray halo, was
required for the impact of the environment to be known.  However,
environmental quenching addresses only half or less of the equation. Galaxies
not only evolve, or rather stagnate, through quenching, but they also grow
through accretions and acquisitions.  Indeed, this dictates the speciation of
galaxies \citep{Graham-triangal}.  It, therefore, seems likely that the galaxy
morphology, a mirror to its merger history, coupled with `environmental
quenching', reflects the actual driving forces of galaxy evolution rather than
`mass quenching' coupled with `environment quenching'
\citep{2010ApJ...721..193P}. The system's mass, which tends to correlate with
the number of major mergers, would then be a secondary, consequential factor.

\citet{2020MNRAS.491...69E} ask, Do bulges stop stars from forming?
Fig.~\ref{Fig8}c plots the sSFR versus the bulge stellar mass.
While there is a sequence from spirals to dust-rich S0s to ES,e and E
galaxies, it is also apparent 
that the non-dusty S0 galaxies with the same bulge masses as the S
galaxies have very different sSFRs.  
This observation is perhaps better illustrated by plotting the sSFR versus the
bulge-to-total stellar mass ratio (Fig.~\ref{Fig7}c). 
It is clear from Fig.~\ref{Fig8}c that the bulge mass does not drive the
star-forming sequence.  Galaxies with the same bulge mass may be a dust-poor
S0 galaxy with no substantial star formation or a spiral galaxy with ample
star formation.

Furthermore, spheroids with a range of stellar
mass from a few billion to a trillion solar masses may be associated with
galaxies having little to no star formation, evidenced by the dust-poor S0 and
E galaxies.  On the other hand, a trend of decreasing sSFR with increasing
galaxy (and spheroid) stellar mass exists for the restricted sample of S, 
dust-rich S0, and E galaxies.  However, as noted, this ignores many intermediate-to-low mass
galaxies, which dominate the galaxy mass function by number, are
disc-dominated, and predominantly dust-poor.

\begin{figure}
\begin{center}
\includegraphics[trim=0.0cm 0cm 0.0cm 0cm, width=1.0\columnwidth, angle=0]{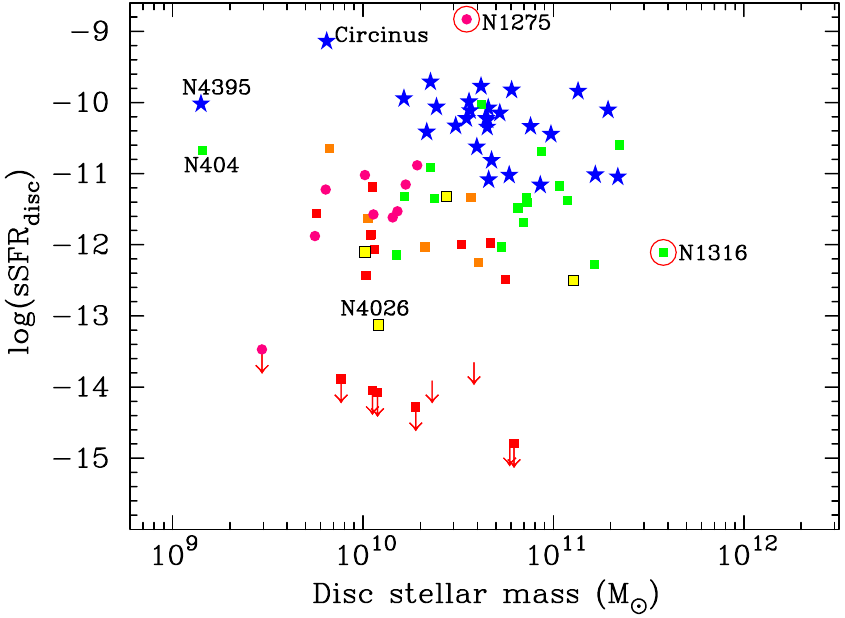}
\caption{Similar to the middle panel of Fig.~\ref{Fig8} but showing the
  {\it disc} sSFR$_{\rm disc}$ versus $M_{\rm *,disc}$. 
}
\label{Fig10}
\end{center}
\end{figure}

\begin{figure}
\begin{center}
\includegraphics[trim=0.0cm 0cm 0.0cm 0cm, width=1.0\columnwidth,
  angle=0]{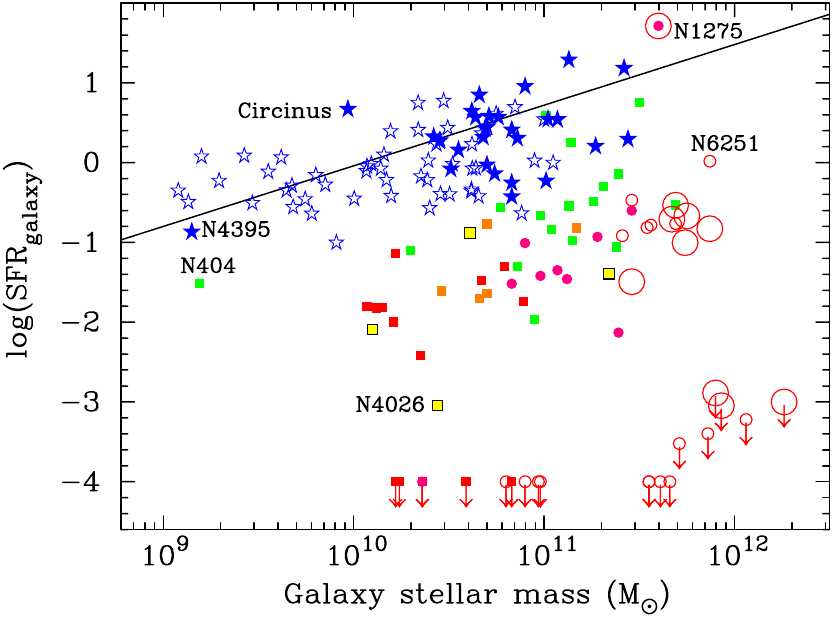}
\caption{
SFR versus galaxy stellar mass.   
The black line is the star-forming main sequence from
\citet{2015ApJ...801L..29R}, without any adjustment for potentially different
galaxy luminosities and stellar mass-to-light ratios between studies.
The dust-rich S0 galaxies, shown in green, are the population which form the
high-mass bridge \citep[aka `green mountain':][]{2018MNRAS.481.1183E} 
between the blue star-forming sequence and the red cloud. 
The Virgo Cluster spiral galaxies from \citet{2022MNRAS.512.3284S} have
been included (open stars). 
As noted in the text, SFR detection limits make it challenging to populate the
lower-left of the diagram 
with the low-mass ETGs that dominate the Universe by number and which are
included in the
\citet{1976ApJ...206..883V} Trident,
the ATLAS$^{3D}$ Comb \citep{2011MNRAS.416.1680C}, 
and the Triangal \citep{Graham-triangal}.
This will be pursued in fututre work.
}
\label{Fig11}
\end{center}
\end{figure}

Fig.~\ref{Fig9} is a variant of 
Fig.~\ref{Fig8}a. 
It shows the $B/T$ flux ratios rather than the morphological
types.  In so doing, it provides an alternative view for the result seen
in
Fig.~\ref{Fig7}c. 
Specifically, it shows that the $B/T$ ratio does not dictate the sSFR of galaxies.  The
morphology of the galaxies --- which better reflects their formation history ---
appears to be a more apt tracer of the sSFR
(Fig.~\ref{Fig8}a). 
Most large surveys have a relatively coarse 
estimate of galaxy morphology, which often results in a rather large bag of
uncertain or intermediate galaxy types. As such, they tend to miss what is
revealed here by using a sample with precise morphologies.
Other complications with large surveys can arise from the necessarily automated bulge/disc 
decompositions of thousands of galaxies. For example, 
many S0 and S galaxies contain substantial bars, and 
the combined flux from the bulge$+$bar can result in a S\'ersic model bulge with an overly large
S\'ersic index, size, and flux for what is supposed to be just the bulge.
In addition, many discs do not have a simple exponential profile, but require a
truncated, anti-truncated, or inclined disc model (not typically used by the
automated routines) in order to obtain the correct $B/T$ ratio. 
Bypassing decompositions, the radial concentration of stars in a galaxy, which
varies 
monotonically with the galaxy's S\'ersic index
\citep{2001MNRAS.326..869T},
is 
sometimes used as a tracer of galaxy type.  In practice, this, too can be problematic for a
couple of reasons.  Measures of concentration using radii thought to contain 80-90 
per cent of the total galaxy flux can be highly sensitive to the image exposure
depth when the S\'ersic index is high 
\citep{2001AJ....122.1707G}.
This can
hinder the separation of pure E galaxies from S0 galaxies with big bulges.  Second,
even when measured correctly, LTGs and low-mass ETGs can have the same low measure
of concentration.  Therefore, the concentration is not always a good discriminant for
separating ETGs from LTGs. 

In the current sample, the S and the dust-rich S0 galaxies have a similar
range of stellar-disc masses, while the S galaxies have the higher sSFRs
(Fig.~\ref{Fig8}b).  
Readers are reminded that the sSFR is typically the SFR
divided by the {\it galaxy}, rather than the {\it disc}, stellar mass.  Suppose the
bulk of new star formation occurs in galactic discs, where gas can condense
and cool.  In that case, it may make more sense to consider trends involving the star
formation rate plotted against the stellar-{\it disc} mass or, equally, the
{\it discs'} specific SFR versus the {\it discs'} stellar mass.  In
Fig.~\ref{Fig10}, we use sSFR$_{\rm disc}$, the SFR normalised by the 
stellar-disc mass.  Due to the dominance of the disc's stellar mass in most disc
galaxies, the trend is qualitatively similar to that seen in the sSFR$_{\rm
  galaxy}$--$M_{\rm *,disc}$ diagram (Fig.~\ref{Fig8}b).  However, as can be
sensed from the distribution of bulge-to-total stellar mass ratios in
Fig.~\ref{Fig7}c, this slightly elevates the sSFR for some S0 galaxies and
considerably so for the ES,e galaxies with their intermediate-scale
discs. Using sSFR$_{\rm disc}$ rather than sSFR$_{\rm galaxy}$ 
reduces the gap seen in Fig.~\ref{Fig8}b.

We next close the loop by returning to the SFR$_{\rm galaxy}$--$M_{\rm 
  *,galaxy}$ diagram (Fig.~\ref{Fig3})  with our full sample now shown in 
Fig.~\ref{Fig11} using the {\it Spitzer}-based stellar masses (Section~\ref{subsec_mass}).
This compares well with \citet[][their figure~1]{2021A&A...649A..39L}. 
The black line in Fig.~\ref{Fig11} shows the relation 
from \citet{2015ApJ...801L..29R} for the star-forming main sequence.
Fig.~\ref{Fig11} also includes the LTGs from 
\citet{2022MNRAS.512.3284S}, better revealing the extent of the star-forming
main sequence than the sample with directly measured BH masses.  
For these Virgo LTGs, the SFR rate has been 
calculated using {\it WISE} photometry and the process described in Section~\ref{Sec_SFR}.
The stellar masses are derived consistently with that used to obtain the {\it
  Spitzer}-based stellar masses for the 
sample with directly measured BH masses. For this additional LTG sample, 
this was done using the 3.6 $\mu$m (AB) galaxy apparent 
magnitudes from the Spitzer Survey of Stellar Structure in Galaxies 
 \citep[S$^4$G:][]{2010PASP..122.1397S} 
catalog\footnote{\url{http://cdsarc.unistra.fr/viz-bin/nph-Cat/html?J/PASP/122/1397/s4g.dat.gz}}. 
The distances from  \citet[][their table~1, column~8]{2022MNRAS.512.3284S} 
 and the following ($B-V$)-dependent (stellar mass)-to-light ratio from  
\citet[][their equation~4]{Graham:Sahu:22a} was also used 
\begin{equation}
\log(M_*/L_{3.6}) = 1.034(B-V) - 1.067. 
\label{Eq_MonL_IP13}
\end{equation}
In total, 56 LTGs with both $B$ and $V$ magnitudes were included.

Fig.~\ref{Fig11} also displays the steep decline in star-formation rate as one progresses
from massive ($\gtrsim$ few $10^{10}$ M$_\odot$) 
spiral galaxies to more massive, dust-rich S0 galaxies and then to E
galaxies and BCGs, which are predominantly E galaxies. This connection led
\citet{2018MNRAS.481.1183E} to advocate for a single sequence connecting the
spiral galaxies with the E galaxies.  
A more complete picture connecting the galaxies involves the triangular-shaped evolutionary
sequence created by \citet{Graham-triangal} and referred to there as the
`Triangal'.  For the first time, in the following section, 
this merger-induced pattern of galaxy growth and metamorphism is
revealed through the sSFR-$M_*$ diagram (Fig.~\ref{Fig12}).

\subsubsection{The Schema}

\begin{figure}
\begin{center}
\includegraphics[trim=0.0cm 0cm 0.0cm 0cm, width=1.0\columnwidth,
  angle=0]{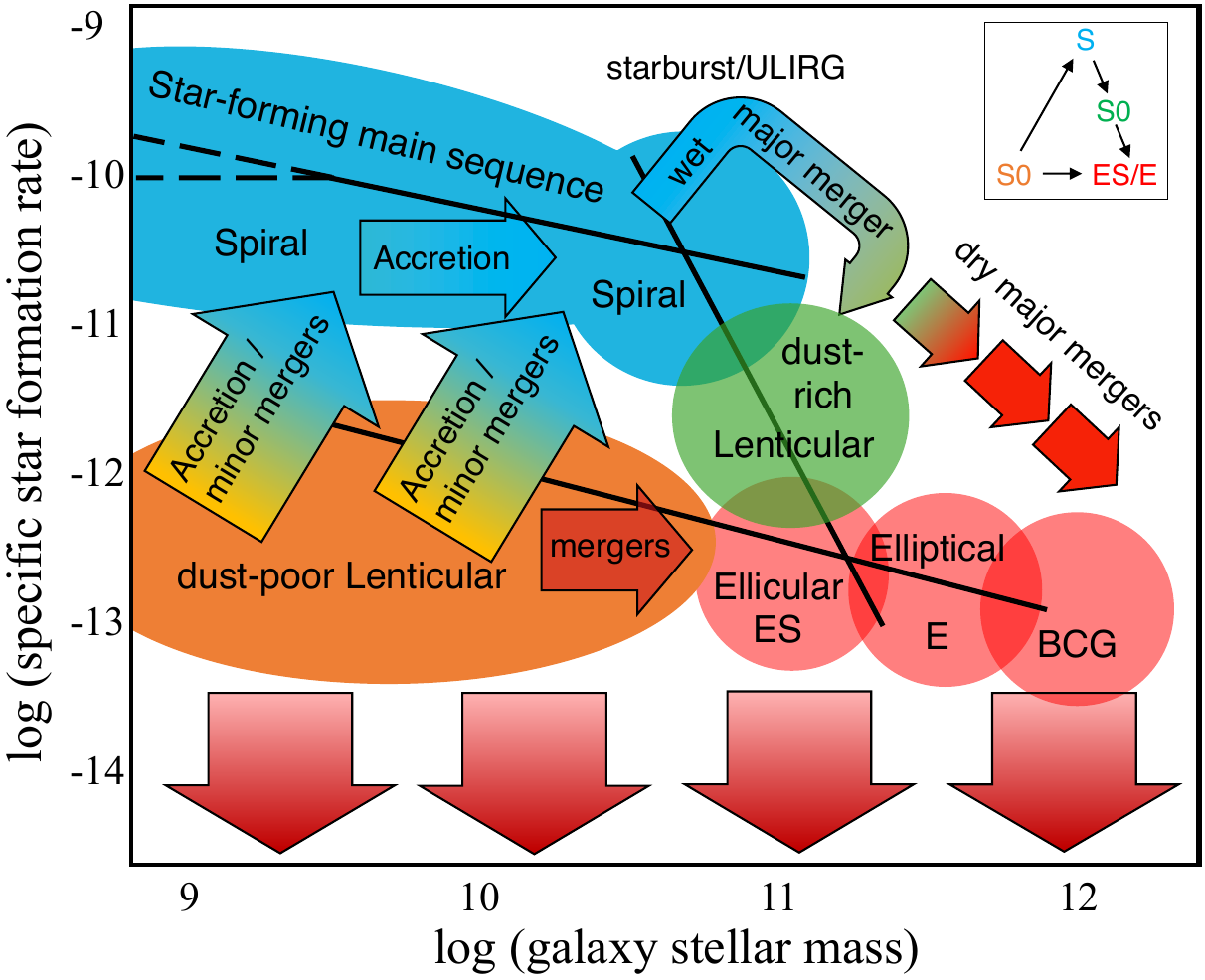}
\caption{ 
Stellar-mass growth sequence from  \citet{Graham-triangal},
tracking the morphological evolution of galaxies. 
\citep[][presents non-growth pathways involving the
rapid cessation of star formation or the removal of stars.]{Graham:Sahu:22b} 
The upper line is the `star-forming main sequence', while 
the lower line 
demarcs the upper envelope of the dust-poor ETGs.  The steep line on the
right tracks the bright galaxy sequence 
\citep[e.g.,][their figure~3]{2018MNRAS.473.3507E} 
as star formation fades after
fuel-consuming merger events and dry mergers. 
The schema shown here captures the left-hand panel of Fig.~\ref{Fig8},
with insight from Fig.~\ref{Fig3}.
Tidal disturbances can send
high-(angular momentum) gas inward, and dwarf/satellite galaxy accretion
can induce spiral arms, thereby, primarily in the younger Universe, elevating
galaxies that were once equivalent to the progenitors of today's dust-poor S0
galaxies onto the star-forming main sequence. 
Most of these S0-to-S transitions likely occurred in the past, while 
(major merger)-induced S-to-S0 transitions are expected to occur in the past, present, and future.
After the potential starbursts from these gas-rich S galaxy collisions fade,
it produces the dust-rich S0 galaxies in/near the `green
valley' of the colour-mass diagram. Merging of these dust-rich S0 
galaxies produces the E galaxies. 
Selection biases are such that the low-mass dust-poor S0 galaxies with low
sSFRs are preferentially missed by surveys seeking star-forming galaxies
\citep[as noted by][their
  figure~3]{2018MNRAS.473.3507E}.
This can partly be rectified with a
V/V$_{\rm max}$ correction \citep[e.g.,][their
  figure~4]{2015ApJ...801L..29R}.
}
\label{Fig12}
\end{center}
\end{figure}

Here we provide an interpretation of the patterns seen in the sSFR-$M_*$ diagram.

As revealed through Fig.~\ref{Fig5}a  and discussed in
\citet{Graham-triangal}, the 
growth of classical bulges and SMBHs 
commences with low-mass and dust-poor S0 galaxies. 
If low-mass galaxies enter the hot X-ray-emitting gas halo of a large galaxy
group or cluster, their reservoir of high-angular momentum cold gas
\citep[e.g.,][]{1989ApJ...347..760C, 1997ApJ...481..157P, 1997AJ....114.2479V,
2023arXiv230905841D} will be ram-pressure stripped, 
along with any dust clouds \citep{1972ApJ...176....1G, 2010AJ....140.1814Y, 2013MNRAS.429.1747M}. 
These galaxies may not have experienced a significant accretion event, gravitational
perturbation, or disc-instability that would
lead them to blossom into an S galaxy.  Due to their cluster-stunting growth, they 
may be primordial galaxies, albeit with an evolved/aged stellar population
\citep{2004AJ....127.1502R, 2004AJ....127.3213M, 2010MNRAS.405..800P, 2013MSAIS..25...93S}. 
Typically having small $B/T$ ratios \citep[][figure~A2]{Graham-triangal}, and thus held-up by rotation, 
they are ETGs.
On the other hand, 
if they are still in the field or a small group environment, they may 
retain substantial H\,{\footnotesize I} fuel reserves, waiting for the right gravitational
disturbance (or capture/merger) to zap the angular momentum of these gas
reserves and send it inward. 
This can reinvigorate star formation \citep{2008MNRAS.385.1903L} and induce a
spiral wave \citep[e.g.,][]{2022MNRAS.516.1114K}, 
as can captured (and internal) gravitational perturbers \citep{2013ApJ...766...34D}, 
taking 
systems onto the spiral galaxy sequence.  This should not be a surprising
origin for the S galaxies; 
in simulations and theories to explain the creation of S galaxies, one
commences with a spiralless disc galaxy, an S0 galaxy.  The S galaxies can be 
built-up from (S0 galaxies via) minor 
mergers \citep{2004ApJ...612..894B}, satellite capture
\citep{2005ApJ...619..807L, 2010AJ....140..962M}, 
and gas accretion via cold flows from
either the cosmic web, disturbed high-angular-momentum gas clouds, or a
cooling hot-gas halo 
\citep{2006MNRAS.368....2D, 2009ApJ...694..396B, 2011MNRAS.415.3750M, 2017MNRAS.467.1083L}.  This increases
their SFR, and recycling material from the new stellar winds produces a 
rolling tail of star formation (in-situ) until the next accretion/merger fuelled burst.

Mergers also bring in new stars (ex-situ) and can redistribute pre-existing
disc stars into the bulge component of a galaxy.  Quantifying this latter
development, coupled with the morphological transition of galaxy type, i.e.,
the speciation of galaxies \citep{Graham-triangal}, could help further advance
some galaxy mass growth models \citep[e.g.][]{2013MNRAS.428.3121M,
  2013MNRAS.435.2445M}.  Rather than treating galaxies as structureless
entities with a single stellar mass --- as done in many surveys and
cosmological and semi-analytical simulations --- a galaxy component analysis
\citep[e.g.,][]{2006MNRAS.371....2A, 2013MNRAS.432.1768K, 2022MNRAS.516..942C}
and morphological awareness\footnote{One area of mismatch 
  that should be noted to avoid confusion is that simulators may refer to S0 galaxies,
  particularly the lower-mass S0 galaxies, as LTGs rather than ETGs
  due to their use of rotation curves and bulge-to-total stellar mass ratios
  for the identification of galaxy type \citep[e.g.,][]{2014MNRAS.444.1518V},
  as opposed to observers who use the presence/absence of spiral structure.
  This differentiation is increasingly relevant at masses less than that of the Milky Way.}  
should aid a fuller understanding of the mass assembly of galaxies
\citep{2006ApJ...651..120B, 2006MNRAS.367.1394K, 
2008ApJ...675..234P, 2013A&A...556A..55I, 2022MNRAS.513..439D} and the size evolution
of discs, bulges and galaxies \citep{2005ApJ...626..680D, 2006ApJ...650...18T, 2014ApJ...788...28V}. Indeed, the benchmark $z\sim0$ size-mass
relation for bulges is quite different to that of galaxies \citep[e.g.,][and
  references therein]{2023MNRAS.519.4651H}.

The broad bend in the star-forming main sequence around $\sim$10$^9$ M$_\odot$
was explored by \citet{2019MNRAS.483.3213P} at $z\sim0$ and 
\citet{2014ApJ...795..104W} over a range of redshifts from 0.5 to 2.5.
Although not strong, a bend appears evident in the ($z\sim0$) S$^4$G data 
(Fig.~\ref{Fig3}). This might be partly due to the presence of low-mass 
S0$\rightarrow$S 
transition objects.  The detection and inclusion of such galaxies reduce 
the sample's median SFR at a fixed stellar mass, shifting the median below the
star-forming main sequence `ridge
line' detected by \citet{2015ApJ...801L..29R} and shown in Fig.~\ref{Fig11}.

\citet{2022MNRAS.513..389R} have suggested that minor mergers are a driving
influence in rejuvenating star formation in S0 galaxies. 
Major mergers can also do this.  Major wet mergers of S galaxies can induce a 
starburst \citep{1997ApJ...478..144S}, build up the bulge using pre-existing
disc stars \citep{2003ApJ...597..893N, 2011MNRAS.416.1654B}, 
and possibly (in the past when more gas was available) 
yield an ultraluminous infrared galaxy \citep[ULIRG:][]{1988ApJ...325...74S} whose 
SFR will decay according to some gas consumption timescale, modulo no further
significant accretion of gas.  The associated enhancement in dust, evident in the
prototypical starburst galaxy M82 \citep{2019PASJ...71...87Y}, 
likely arises from the condensation of metals as the metal-enriched gas cools to form a
new generation of stars \citep{2004ApJ...606..862O, 
2021MNRAS.504...53S}.  As such, the dust-to-gas ratio may be
elevated in dust-rich S0 galaxies, and their gas-to-stellar mass ratio is reduced
in the evolved dust-rich S0 galaxies. That is, the notion of a near-constant
dust-to-gas ratio is not expected to apply across the spectrum of
disc-dominated galaxies. Indeed, the dust-to-H{\footnotesize I} gas ratio 
increases with metallicity, [O/H], \citep[][their figure~6]{2008ApJ...678..804E}.
This dust and metal build-up in galaxies with increasing stellar mass
\citep[e.g.,][]{2002MNRAS.330..876C, 2011MNRAS.417..785J} works in tandem with galactic winds and
ram-pressure stripping, which operate more effectively at lower masses 
\citep[e.g.][]{2004ApJ...613..898T, 2006MNRAS.369.1021M}.

A major merger of two dust-rich S0 galaxies erodes the stellar discs. It 
yields a massive elliptical galaxy or an ES galaxy if some remnant disc
remains.  Decades ago, it was speculated that
colliding S galaxies would generate a hot mess, i.e., an elliptical galaxy
\citep{1977egsp.conf..401T}.  However, subsequent modellers pointed out that 
unless there are specially contrived orbital encounters, it 
typically takes more than one major merger for the net rotation, and thus
discs, to entirely cancel \citep[e.g.,][]{1996ApJ...471..115B, 2003ApJ...597..893N}.
A major collision of sufficiently massive dust-poor S0 galaxies may bypass the
S galaxy phase to create an S0 with a more massive spheroid.  The dusty nature
of the massive S0 galaxies suggests their creation involved wet rather
than dry mergers, perhaps occurring in a much younger Universe (when today's
dust-poor S0 galaxies were gas-rich).  

The chain of events noted above is shown in Fig.~\ref{Fig12}.  The schematic
seen there is a direct mapping of the quasi-triangular-shaped galaxy anatomy
sequence, with evolutionary pathways, introduced in \citet{Graham-triangal}.
Although not explored here, one would expect clumpiness and asymmetry to
broadly track the morphological types shown in
Fig.~\ref{Fig12},
from smooth
and symmetric (dust-poor S0 galaxies) to the opposite appearance (S galaxies,
starbursts) to a middle ground (dust-rich S0 and some ES galaxies) before
settling back down to a more smooth and relaxed appearance (E galaxies).
As suggested in \citet{Graham-triangal}, the lower-limit to the dust-poor S0
galaxies may extend into the realm of the
 low central surface brightness ETGs \citet{1984AJ.....89..919S},  
nowadays referred to as ultra diffuse galaxies \citep[aka
  UDGs][]{2015ApJ...798L..45V}.

\section{Summary} 
\label{Summ}

Armed with the (galaxy morphology)-dependent $M_{\rm
  bh}$-$M_{\rm *,sph}$ sequences (Fig.~\ref{Fig5}a), other trends in this diagram can
be better understood, such as the distribution of star formation rates (Fig.~\ref{Fig5}b).
For instance, after identifying the spiral and dust-rich lenticular galaxies, 
it is apparent they are bookended by low-star-formation (dust-poor S0 and E)
galaxies.  Without substantial numbers of low-mass S0 galaxies, and excluding
mergers like NGC~5128 (Cen~A), something of an over-arching band (rather than
bookends) of low-SFR galaxies appears above the S galaxies in this diagram
\citep{2016ApJ...817...21S, 2017ApJ...844..170T}.  That is, with a restricted
sample, quiescent galaxies appear to have more massive black holes than
star-forming galaxies at a given stellar mass, suggesting a role for AGN
feedback, with higher $M_{\rm bh}/M_{\rm *}$ ratios quenching star formation.

However, this is an incomplete picture.  Fig.~\ref{Fig7} reveals that the galaxy
morphology is a better indicator of the sSFR than the $M_{\rm bh}/M_{\rm *}$
ratio. It is also a better indicator than the prominence of the bulge,
quantified by the $M_{\rm *,sph}/M_{\rm *,gal}$ ratio.  We additionally show
that the bulge mass does not determine the occurrence on/off the star-forming
main sequence.
Galaxies with higher SFRs primarily have these because of refuelling events
which have dictated their morphology.  They are foremost spiral galaxies, or
dust-rich S0 galaxies built via major wet mergers, rather than systems with a
low specific black hole mass, i.e., a lower $M_{\rm bh}/M_{\rm *}$ ratio that
is less capable of quenching star formation.

Some dust-rich S0 galaxies fill 
the `green valley', forming a bridge or `green mountain'
\citep{2018MNRAS.481.1183E} between the blue
star-forming main sequence and the `red cloud' of E galaxies in the
(s)SFR-$M_{\rm *,gal}$ diagram 
(Fig.~\ref{Fig3} and \ref{Fig8}).
This is not due to AGN (Fig.~\ref{Fig4}) and is also evident when using
the stellar mass of the disc (Fig.~\ref{Fig10}), where the bulk of the star
formation occurs.  
The `red cloud' in the (s)SFR-$M_{\rm *,gal}$ diagram is seen to become a more extended  
`red sequence' (or more correctly, an upper envelope) upon the inclusion of 
dust-poor S0 galaxies.  
 
Fig.~\ref{Fig12} provides a schematic representation of evolutionary pathways
through the sSFR$_{\rm galaxy}$-$M_{\rm *,galaxy}$. 
It is an adaption of the morphological transitions seen
in Fig.~\ref{Fig5} and previously captured by the `Triangal' for understanding
the speciation of galaxies \citep{Graham-triangal}.

\section*{Acknowledgements}

T.H.J.\ is grateful for an extended visit to Swinburne University of
Technology and acknowledges support from the South African National Research
Foundation.
This publication uses data products from the Wide-field Infrared
Survey Explorer, a joint project of the University of California, Los
Angeles, and the Jet Propulsion Laboratory/California Institute of Technology,
funded by the National Aeronautics and Space Administration (NASA).
This research has also made use of NASA's Astrophysics Data System Bibliographic
Services and the NASA/IPAC Extragalactic Database (NED),
which is funded by NASA and operated by the California Institute of Technology.

\section{Data Availability}

Black hole masses, along with {\it SST}-derived  
spheroid and galaxy stellar masses, are tabulated
in \citet{Graham:Sahu:22a}. 
The spheroid's stellar masses are obtained from published 
multicomponent decompositions, which separate bars and inner discs from the
spheroids \citep{2016ApJS..222...10S, 2019ApJ...873...85D,
  2019ApJ...876..155S, Graham:Sahu:22b}. 
Galaxy morphologies --- including whether the E galaxies are BCG or
cD, and the `dust bin' of each S0 galaxy --- are provided in \citet{Graham:Sahu:22a}
and \citet{Graham:Sahu:22b}, 
and \citep{Graham-S0}, respectively.

% \bibliographystyle{mn2e}
\bibliographystyle{mnras}
\bibliography{Paper-BH-SFR}{}

\appendix

% \clearpage 
% 
% \newpage

\section{Notes on individual galaxies}
\label{Apdx1}

Some galaxies were labelled in the figures or noted in passing because they
stood out for one reason or another and may, therefore, be of interest.  Below
are some quick comments on these galaxies.

Circinus has a significant Seyfert AGN \citep{2022A&A...664A.142T,
  2023MNRAS.519.3237S}; thus, the {\it WISE}-determined sSFR is not
applicable/correct.

NGC~1275 is the Perseus cluster's BCG, with a substantial AGN
\citep{1965ApJ...142.1351B, 2001MNRAS.328..359I}.  The {\it WISE}-determined
sSFR is again not applicable/correct.

NGC~1300 is an outlying S galaxy in the $M_{\rm bh}$-$M_{\rm *,sph}$ diagram
and was excluded from the Bayesian analyses for the S galaxies
\citep{Graham-triangal}.  It has a high S\'ersic index in the $M_{\rm
  *,sph}$--(S\'ersic, $n$) diagram \citep[their Fig.~2]{2020ApJ...903...97S}
and may have an unmodelled barlens component
\citep[][footnote~8]{2023MNRAS.518.6293G}, but this is unlikely to increase
the spheroid mass.

NGC~1316 is the Fornax cluster's BCG.  While many/most BCGs are E galaxies,
this is a dusty S0 galaxy, likely built, in part, through the acquisition of a
spiral galaxy \citep{2004ApJ...613L.121G, 2021JApA...42...34V}.

NGC~1332 is a relatively dust-free, relic compact massive galaxy without a
large-scale disc \citep{2016MNRAS.457..320S}.  It is designated an ES,b
galaxy, midway between the dust-poor S0 and E galaxies.  It resides in a
region of the $M_{\rm bh}$-$M_{\rm *,sph}$ diagram where one finds the bulges
of dusty S0 galaxies.

NGC~2787 is a reasonably dusty S0 galaxy located to the left of the $M_{\rm
  bh}$-$M_{\rm *,sph}$ relation for dust-poor S0 galaxies. On one hemisphere,
it has strong dust rings over the inner 160 pc, and weaker dust rings out to
400 pc.  It has a pseudobulge, a classical bulge \citep{2003ApJ...597..929E},
a polar disc and nuclear rings that do not reside in the primary disc plane
\citep{2004AJ....127.2641S}.

NGC~3489 is a dust-rich S0 galaxy located on the low-mass end of the dust-poor
S0 sequence in the $M_{\rm bh}$-$M_{\rm *,sph}$ diagram.  The measurements are
not thought to be in error; instead, some outlying systems may reflect that
the trends are not perfect laws of physics.

NGC~4026 comprises an old stellar population and contains no significant
H{\footnotesize I} gas of its own.  However, it is the dominant galaxy in its
small group and has accreted a $10^8$ M$_\odot$ filament of H{\footnotesize I}
gas \citep{2003ApJ...584..260W}.

NGC~4342 is a tidally-stripped galaxy whose spheroid mass has been reduced by
an unknown amount \citep{2014MNRAS.439.2420B}.

NGC~6251 has a Seyfert 2 AGN, a dusty nuclear disc, and a strong one-sided
radio jet \citep{1984ApJS...54..291P, 1997ApJ...486L..91C}.  It resides in the
$M_{\rm bh}$-$M_{\rm *,sph}$ diagram, where one would expect to find BCGs.  It
is an E galaxy likely built from multiple major mergers but is not a BCG.

\section{Further commentary on AGN}
\label{Apdx2}

There are a couple of notable exceptions in Fig.~\ref{Fig4} that warrant
mention, in that they contain AGN but do not appear in the QSO/AGN box shown
there. 
The spectacular radio galaxy, NGC~5128 \citep[Centaurus~A:][]{1999A&A...341..667M}, has a very bright
nucleus, 
graced by a radio-jet and Mpc-sized non-thermal radio plume, and is one of the
closest 
AGNs.  However, the host galaxy is massive and extended and contains a dusty,
star-forming remnant of 
a past 'wet' merger (see Fig.~\ref{Fig2}). Consequently, the {\it WISE}
colours
are perfectly consistent with normal thermal emission, even though it 
is centred around a powerful non-thermal engine.  In general,
it has been found that galaxies
may harbour a radio-quiet AGN, which is revealed, e.g., with optical
spectroscopy via
line ratios, 
but the {\it WISE} colours may only be sensitive to the greater host (disc $+$ bulge
populations).
\citet{2022ApJ...939...26Y} noted that 
if one is to comprehensively reveal AGNs in nearby galaxies, it requires
a complete multi-wavelength approach, including the X-ray (nuclear activity and
jets), optical (kinematics and line ratios), 
infrared (dust-obscured AGNs), and radio (non-thermal emission,
jets and plumes).
The upshot is that NGC~5128, along with NGC~1275, have nuclear (central) W1$-$W2 colours around
1.0, while the global or total colour is $\sim$0.0 and $\sim$0.1, respectively.

Among our full sample, we recognise an additional five (along with NGC~1194
and NGC~5252)  `warm AGN' whose
SFRs will be over-estimated to some extent. They are the spiral galaxies 
Circinus, NGC~1320, NGC~4151, NGC~4388 and NGC~7582. 
An additional three galaxies have $0.25<(W1-W2)<0.4$.  They are the spiral
galaxies NGC~2273 and NGC~3227, along with the Perseus cluster's BCG NGC~1275,
aka
Perseus~A, which is an ES (ellicular) galaxy.  
As noted above, this type 1.5 Seyfert galaxy is initially excluded from some
figures, as is Circinus, for clarity.  We note that as with NGC~5128, these
``warm'' 
galaxies have colours dominated by starlight, with only a small 
contribution from the AGN.   When, however, we consider only the colour of the
nuclear region (using an aperture set by the angular resolution of the {\it WISE}
imaging), 
the resulting W1-W2 colour can be well above a value of unity, residing in the
AGN/QSO region of the colour-colour diagram.

Galaxies with
more captured cool gas are expected to have higher SFRs and greater AGN
activity. However, as noted above, we are mindful that AGN-heated dust may
interfere with some {\it WISE}-estimated SFRs. 
Focussing on the 40 S0 galaxies identified in  \citet{Graham-S0},
five stand out in the {\it WISE} images 
for their high (non-stellar)-to-stellar mid-infrared emission level.
They are the low-ionization nuclear 
emission-line region (LINER) dwarf galaxy 
NGC~404 \citep[Seyfert:][]{1985ApJS...57..503F, 1985ApJ...299..443W,
  2006A&A...455..773V}, 
NGC~1194 \citep[dusty Seyfert 1.9:][]{1988ApJ...326..653K,
  2006A&A...455..773V, 2015AJ....149..155K}, 
NGC~4594 \citep[Sombrero galaxy with large dust disc and Seyfert 2 nucleus:][]
    {1985A&A...143..334B, 1985ApJS...57..503F, 2002A&A...394..435P}, 
NGC~5128 \citep[Centaurus~A, an unrelaxed merger with a radio
  jet:][]{1998A&ARv...8..237I}, and 
NGC~5252 \citep[relic quasar / LINER / dusty Seyfert
  1.9:][]{1982AJ.....87.1628H, 1998A&A...333..877G, 2005A&A...431..465C,
  2015AJ....149..155K}. 
NGC~1194 and NGC~5252 have both been flagged above for containing a warm AGN,
and they have been flagged in the literature for potentially hosting a binary
black hole \citep{2015ApJ...814....8K, 2017MNRAS.464L..70Y,
  2015KPCB...31...13V, 2016AN....337...96F}.  This is unsurprising given that
the higher-mass dust-rich S0 galaxies have been built from gas-rich major
mergers \citep[][table~2]{Graham-S0}.

In addition to the above five galaxies, 
the strong dust lanes in the merger remnants NGC~1316 (also with a radio jet)
and
NGC~5018 stand out in the {\it WISE} images due to their red MIR colour. 
The only other dust-rich S0 galaxy with an SFR greater than 0.5 M$_\odot$
yr$^{-1}$ 
is NGC~3665, and it, too, has a radio jet \citep{1984ARA&A..22..319B,
  1992A&AS...95..249L, 2002A&A...381..757L}. 
These eight galaxies have been marked in Fig.~\ref{Fig3}, and the first
six are shown in Fig.~\ref{Fig2}. 

The above eight S0 galaxies are dust-rich (dust$=$Y).
Eight of the nine remaining dust-rich S0 galaxies do not have active nuclei. 
The exception is 
NGC 2974 \citep[Seyfert 2:][]{2003AJ....126.1750M, 2006A&A...455..773V}. 

Among the four `dust$=$y' S0 galaxies, two contain an AGN. 
They are NGC~2787 and NGC~3998, which contain a 
LINER with broad Balmer lines \citep{1997ApJS..112..391H,
  2006A&A...455..773V}. 
In contrast, none of the 19 `dust$=$n' or `dust$=$N' S0 galaxies 
contain an AGN according to the Activity Type in
the NASA/IPAC Extragalactic Database
(NED)\footnote{\url{http://nedwww.ipac.caltech.edu}}.
This includes five `dust$=$n' S0 galaxies, revealing that their dusty nuclear
discs/rings are not powering an AGN.  

For reference, among the 
28 S galaxies (including Circinus), 21 contain a (predominantly Seyfert) AGN. 
In contrast, 
among the 35 remaining ETGs, 
5/9 BCG (excluding the already mentioned S0 BCG NGC~1316 with its radio jet),
2/9 non-BCG ES,e, 
and 6/17 E 
contain an AGN.
These 13 AGN, in 35 ETGs, consist of five Seyfert~2 nuclei, two LINERs (one
with broad
Balmer lines, and six Flat-Spectrum Radio Sources (FSRSs). 

Another way to dice things is to report that there are 
3/28 S galaxies with a radio jet or FSRS, 
3/39 non-BCG S0 with a radio jet or FSRS, and 
7/36 ETGs (17 E $+$ 9 ES,e $+$ 10 BCG) with a radio jet or FSRS.
This abundance ratio 
roughly represents 10 per cent for the two former galaxy types and 20 per cent
for the latter.

\section{Specific star formation rates}
\label{Apdx3}

\begin{table*}
\caption{Star formation rates}\label{Table-data}
\begin{tabular}{llccccc}
\hline
Galaxy   &  Type     & Dust &    W1$-$W2   &     W2$-$W3 & $\log(M_{\rm *,gal}/{\rm M}_\odot)$ & SFR \\
         &           &      &    mag       &      mag    &     dex & M$_\odot$ yr$^{-1}$  \\
\hline
Circinus & S          &   &  0.69$\pm$0.03 &  4.02$\pm$0.03 &  10.04$\pm$0.09 &  4.6493$\pm$0.9689 \\ 
IC 1459  & E          &   & -0.05$\pm$0.04 &  0.39$\pm$0.06 &  11.24$\pm$0.08 &  0.1724$\pm$0.0262 \\ 
IC 2560  & S          &   &  0.19$\pm$0.04 &  3.38$\pm$0.04 &  10.43$\pm$0.08 &  2.6410$\pm$0.2648 \\ 
IC 4296  & BCG (E)    &   & -0.08$\pm$0.04 &  0.02$\pm$0.08 &  11.47$\pm$0.08 &  0.0994$\pm$0.0181 \\ 
NGC 0224 & S          &   & -0.03$\pm$0.03 &  2.08$\pm$0.04 &  10.71$\pm$0.08 &  0.5871$\pm$0.0568 \\ 
NGC 0253 & S          &   &  0.20$\pm$0.03 &  3.81$\pm$0.04 &  10.43$\pm$0.08 &  3.7837$\pm$0.3807 \\
NGC 0404 & S0         & Y &  0.03$\pm$0.04 &  1.28$\pm$0.05 &  08.85$\pm$0.09 &  0.0306$\pm$0.0092 \\ 
NGC 0524 & S0         & Y & -0.02$\pm$0.04 &  0.52$\pm$0.06 &  11.10$\pm$0.08 &  0.0871$\pm$0.0112 \\
NGC 0821 & ES,e       &   & -0.04$\pm$0.04 &  0.27$\pm$0.13 &  10.64$\pm$0.08 &  0.0982$\pm$0.0166 \\ 
NGC 1023 & S0         & N & -0.04$\pm$0.03 &  0.18$\pm$0.07 &  10.61$\pm$0.08 &  0.0182$\pm$0.0031 \\
NGC 1097 & S          &   &  0.10$\pm$0.04 &  3.41$\pm$0.04 &  11.22$\pm$0.08 & 15.1384$\pm$2.1432 \\ 
NGC 1194 & S0         & Y &  0.89$\pm$0.03 &  2.83$\pm$0.04 &  10.46$\pm$0.08 &  3.9039$\pm$0.3943 \\ 
NGC 1275 & BCG (ES,e) &   &  0.32$\pm$0.04 &  3.00$\pm$0.04 &  11.52$\pm$0.09 & 51.7151$\pm$8.1149 \\ 
NGC 1300 & S          &   &  0.05$\pm$0.04 &  2.91$\pm$0.04 &  10.56$\pm$0.08 &  1.4286$\pm$0.1412 \\ 
NGC 1316 & BCG (S0)   & Y & -0.03$\pm$0.04 &  0.65$\pm$0.05 &  11.43$\pm$0.08 &  0.2939$\pm$0.0319 \\ 
NGC 1320 & S          &   &  0.64$\pm$0.04 &  3.34$\pm$0.04 &  10.13$\pm$0.09 &  2.7987$\pm$0.2803 \\ 
NGC 1332 & S0 (ES,b)  & n & -0.03$\pm$0.04 &  0.42$\pm$0.05 &  10.88$\pm$0.08 &  0.1514$\pm$0.0243 \\
NGC 1374 & S0         & N & -0.06$\pm$0.04 &  0.12$\pm$0.07 &  10.33$\pm$0.08 & $<$0.0001 \\	 
NGC 1398 & S          &   & -0.00$\pm$0.04 &  2.14$\pm$0.04 &  11.17$\pm$0.08 &  1.9506$\pm$0.1975 \\
NGC 1399 & BCG (E)    &   & -0.08$\pm$0.04 &  0.17$\pm$0.07 &  11.23$\pm$0.08 &  0.0321$\pm$0.0073 \\ 
NGC 1407 & E          &   & -0.08$\pm$0.04 &  0.07$\pm$0.09 &  11.39$\pm$0.08 & $<$0.0001 \\	 
NGC 1600 & E          &   & -0.09$\pm$0.04 & -0.34$\pm$0.10 &  11.71$\pm$0.09 & $<$0.0006 \\	 
NGC 2273 & S          &   &  0.29$\pm$0.03 &  3.14$\pm$0.04 &  10.43$\pm$0.08 &  2.5597$\pm$0.2557 \\ 
NGC 2549 & S0         & N & -0.05$\pm$0.03 &  0.33$\pm$0.06 &  09.97$\pm$0.08 &  0.0100$\pm$0.0029 \\
NGC 2778 & S0         & N & -0.06$\pm$0.04 &  0.12$\pm$0.05 &  09.89$\pm$0.08 &  0.0154$\pm$0.0052 \\
NGC 2787 & S0         & y & -0.03$\pm$0.03 &  0.59$\pm$0.04 &  09.80$\pm$0.08 &  0.0081$\pm$0.0021 \\ 
NGC 2960 & S          &   &  0.06$\pm$0.04 &  2.98$\pm$0.04 &  10.72$\pm$0.08 &  2.0085$\pm$0.2022 \\ 
NGC 2974 & S0         & Y & -0.04$\pm$0.04 &  1.36$\pm$0.08 &  10.61$\pm$0.08 &  0.2160$\pm$0.0240 \\ 
NGC 3031 & S          &   & -0.03$\pm$0.03 &  1.80$\pm$0.03 &  10.57$\pm$0.08 &  0.3724$\pm$0.0356 \\ 
NGC 3079 & S          &   &  0.24$\pm$0.03 &  3.64$\pm$0.04 &  10.41$\pm$0.08 &  3.6759$\pm$0.3683 \\ 
NGC 3091 & E          &   & -0.03$\pm$0.04 & -0.21$\pm$0.11 &  11.39$\pm$0.09 & $<$0.0004 \\	 
NGC 3115 & ES,b       & Y & -0.01$\pm$0.04 &  0.14$\pm$0.12 &  10.63$\pm$0.08 &  0.0108$\pm$0.0025 \\
NGC 3227 & S          &   &  0.28$\pm$0.04 &  3.09$\pm$0.04 &  10.65$\pm$0.08 &  3.4491$\pm$0.3488 \\ 
NGC 3245 & S0         & n & -0.03$\pm$0.04 &  1.09$\pm$0.04 &  10.45$\pm$0.08 &  0.1691$\pm$0.0163 \\ 
NGC 3368 & S          &   &  0.04$\pm$0.04 &  2.13$\pm$0.04 &  10.55$\pm$0.08 &  0.5551$\pm$0.0541 \\
NGC 3377 & ES,e       &   & -0.05$\pm$0.04 & -0.09$\pm$0.08 &  10.13$\pm$0.08 & $<$0.0001 \\	 
NGC 3379 & E          &   & -0.08$\pm$0.04 &  0.12$\pm$0.05 &  10.64$\pm$0.08 & $<$0.0001 \\	 
NGC 3384 & S0         & N & -0.04$\pm$0.04 &  0.24$\pm$0.05 &  10.37$\pm$0.08 &  0.0334$\pm$0.0054 \\
NGC 3414 & ES,e       &   & -0.04$\pm$0.04 &  0.50$\pm$0.07 &  10.63$\pm$0.08 &  0.0381$\pm$0.0049 \\
NGC 3489 & S0         & Y &  0.01$\pm$0.04 &  1.16$\pm$0.04 &  10.14$\pm$0.08 &  0.0795$\pm$0.0075 \\
NGC 3585 & ES,e       &   & -0.05$\pm$0.04 &  0.06$\pm$0.09 &  11.03$\pm$0.09 &  0.0074$\pm$0.0045 \\
NGC 3607 & ES,e       &   & -0.08$\pm$0.04 &  0.73$\pm$0.06 &  11.13$\pm$0.08 &  0.2514$\pm$0.0261 \\ 
NGC 3608 & E          &   & -0.04$\pm$0.04 & -0.13$\pm$0.10 &  10.56$\pm$0.09 & $<$0.0001 \\	 
NGC 3627 & S          &   &  0.11$\pm$0.04 &  3.44$\pm$0.04 &  10.63$\pm$0.08 &  3.6944$\pm$0.3702 \\ 
NGC 3665 & S0         & Y & -0.05$\pm$0.04 &  1.33$\pm$0.04 &  11.13$\pm$0.08 &  0.7200$\pm$0.0709 \\ 
NGC 3842 & BCG (E)    &   & -0.08$\pm$0.06 & -0.43$\pm$0.07 &  11.45$\pm$0.09 & $<$0.0009 \\	 
NGC 3923 & E          &   & -0.10$\pm$0.04 & -0.03$\pm$0.08 &  11.30$\pm$0.08 & $<$0.0001 \\	 
NGC 3998 & S0         & y & -0.00$\pm$0.04 &  1.39$\pm$0.05 &  10.30$\pm$0.08 &  0.1304$\pm$0.0127 \\ 
NGC 4026 & S0         & y & -0.04$\pm$0.03 &  0.52$\pm$0.05 &  10.18$\pm$0.08 &  0.0009$\pm$0.0008 \\ 
NGC 4151 & S          &   &  0.79$\pm$0.04 &  2.82$\pm$0.04 &  10.53$\pm$0.08 &  4.3674$\pm$0.4406 \\ 
NGC 4258 & S          &   &  0.04$\pm$0.04 &  2.44$\pm$0.04 &  10.56$\pm$0.08 &  0.9339$\pm$0.0918 \\ 
NGC 4261 & E          &   & -0.08$\pm$0.04 &  0.22$\pm$0.08 &  11.17$\pm$0.08 &  0.1530$\pm$0.0231 \\ 
NGC 4291 & E          &   &  0.01$\pm$0.04 &  0.00$\pm$0.09 &  10.47$\pm$0.08 & $<$0.0001 \\	 
NGC 4303 & S          &   &  0.15$\pm$0.04 &  3.87$\pm$0.04 &  10.71$\pm$0.09 &  7.0129$\pm$0.8237 \\ 
NGC 4339 & S0         & N & -0.08$\pm$0.04 &  0.67$\pm$0.14 &  10.02$\pm$0.08 &  0.0728$\pm$0.0160 \\
NGC 4342 & S0         & N & -0.07$\pm$0.04 &  0.31$\pm$0.04 &  10.10$\pm$0.08 & $<$0.0001 \\	 
NGC 4350 & S0         & n & -0.06$\pm$0.04 &  0.51$\pm$0.06 &  10.35$\pm$0.08 &  0.0198$\pm$0.0028 \\
NGC 4371 & S0         & n & -0.03$\pm$0.04 &  0.64$\pm$0.09 &  10.38$\pm$0.08 &  0.0227$\pm$0.0041 \\
NGC 4374 & E          &   & -0.03$\pm$0.04 & -0.04$\pm$0.07 &  11.14$\pm$0.08 & $<$0.0001 \\	 
NGC 4388 & S          &   &  0.43$\pm$0.04 &  3.15$\pm$0.04 &  10.12$\pm$0.08 &  1.8543$\pm$0.1848 \\ 
NGC 4395 & S          &   &  0.08$\pm$0.07 &  2.20$\pm$0.12 &  09.03$\pm$0.09 &  0.1345$\pm$0.0347 \\ 
NGC 4429 & S0         & Y & -0.04$\pm$0.04 &  0.86$\pm$0.05 &  10.75$\pm$0.08 &  0.1446$\pm$0.0146 \\ 
NGC 4434 & S0         & N & -0.04$\pm$0.04 & -0.00$\pm$0.06 &  10.03$\pm$0.08 & $<$0.0001          \\     
\hline
\end{tabular}
\end{table*}

\setcounter{table}{0}

\begin{table*}
\caption{Continued}
\begin{tabular}{llccccc}
\hline
Galaxy   &  Type     & Dust &    W1$-$W2   &     W2$-$W3 & $\log(M_{\rm *,gal}/{\rm M}_\odot)$ & SFR \\
         &           &      &    mag       &      mag    &     dex & M$_\odot$ yr$^{-1}$  \\
\hline
NGC 4459 & S0         & Y & -0.02$\pm$0.04 &  1.17$\pm$0.10 &  10.56$\pm$0.08 &  0.2756$\pm$0.0299 \\ 
NGC 4472 & BCG (E)    &   & -0.09$\pm$0.04 &  0.28$\pm$0.12 &  11.41$\pm$0.08 &  0.2140$\pm$0.0271 \\ 
NGC 4473 & ES,e       &   & -0.06$\pm$0.04 &  0.15$\pm$0.08 &  10.53$\pm$0.08 &  0.0303$\pm$0.0049 \\
NGC 4486 & BCG (E)    &   & -0.07$\pm$0.04 &  0.33$\pm$0.05 &  11.31$\pm$0.08 &  0.1950$\pm$0.0294 \\ 
NGC 4501 & S          &   &  0.06$\pm$0.04 &  3.05$\pm$0.04 &  10.89$\pm$0.08 &  3.4779$\pm$0.3495 \\ 
NGC 4526 & S0         & Y & -0.03$\pm$0.04 &  1.14$\pm$0.05 &  10.84$\pm$0.08 &  0.2872$\pm$0.0281 \\ 
NGC 4552 & ES,e       &   & -0.07$\pm$0.04 &  0.50$\pm$0.10 &  10.77$\pm$0.09 &  0.0448$\pm$0.0065 \\ 
NGC 4564 & S0         & N & -0.04$\pm$0.04 &  0.31$\pm$0.05 &  10.12$\pm$0.08 &  0.0038$\pm$0.0009 \\ 
NGC 4578 & S0         & N & -0.07$\pm$0.04 & -0.24$\pm$0.10 &  10.05$\pm$0.08 & $<$0.0001 \\	 
NGC 4594 & S0         & Y & -0.02$\pm$0.04 &  0.90$\pm$0.05 &  11.06$\pm$0.08 &  0.3270$\pm$0.0325 \\ 
NGC 4596 & S0         & Y & -0.00$\pm$0.04 &  0.34$\pm$0.08 &  10.54$\pm$0.08 &  0.0500$\pm$0.0057 \\
NGC 4621 & ES,e       &   & -0.09$\pm$0.04 &  0.42$\pm$0.13 &  10.89$\pm$0.08 &  0.1176$\pm$0.0146 \\
NGC 4649 & E          &   & -0.07$\pm$0.04 &  0.42$\pm$0.10 &  11.24$\pm$0.08 &  0.1649$\pm$0.0190 \\
NGC 4697 & ES,e       &   & -0.05$\pm$0.04 &  0.09$\pm$0.06 &  10.65$\pm$0.08 &  0.0347$\pm$0.0057 \\
NGC 4699 & S          &   & -0.00$\pm$0.04 &  2.20$\pm$0.04 &  11.06$\pm$0.08 &  1.5938$\pm$0.1583 \\
NGC 4736 & S          &   &  0.02$\pm$0.04 &  2.71$\pm$0.04 &  10.38$\pm$0.08 &  0.8263$\pm$0.0802 \\ 
NGC 4742 & S0         & N & -0.02$\pm$0.04 &  0.40$\pm$0.05 &  09.99$\pm$0.08 &  0.0158$\pm$0.0044 \\
NGC 4762 & S0         & N & -0.04$\pm$0.04 &  0.18$\pm$0.07 &  10.56$\pm$0.08 & $<$0.0001 \\	 
NGC 4826 & S          &   &  0.02$\pm$0.04 &  2.21$\pm$0.04 &  10.54$\pm$0.08 &  0.7212$\pm$0.0698 \\ 
NGC 4889 & BCG (E)    &   & -0.07$\pm$0.04 & -0.17$\pm$0.09 &  11.72$\pm$0.09 & $<$0.0010 \\	 
NGC 4945 & S          &   &  0.20$\pm$0.03 &  3.56$\pm$0.03 &  10.23$\pm$0.08 &  2.0911$\pm$0.2064 \\ 
NGC 5018 & S0         & Y &  0.03$\pm$0.04 &  0.89$\pm$0.07 &  11.10$\pm$0.08 &  0.4968$\pm$0.0514 \\
NGC 5077 & E          &   & -0.06$\pm$0.04 &  0.22$\pm$0.07 &  11.02$\pm$0.08 &  0.1213$\pm$0.0188 \\ 
NGC 5128 & S0         & Y &  0.02$\pm$0.03 &  2.53$\pm$0.03 &  10.86$\pm$0.08 &  1.7869$\pm$0.1759 \\ 
NGC 5252 & S0         & Y &  0.58$\pm$0.04 &  2.27$\pm$0.04 &  11.05$\pm$0.08 &  5.6485$\pm$0.6099 \\ 
NGC 5419 & BCG (E)    &   & -0.09$\pm$0.04 &  0.04$\pm$0.12 &  11.64$\pm$0.08 &  0.1488$\pm$0.0284 \\ 
NGC 5576 & E          &   & -0.03$\pm$0.04 & -0.23$\pm$0.05 &  10.70$\pm$0.08 & $<$0.0001 \\	 
NGC 5813 & S0         & y & -0.04$\pm$0.04 &  0.03$\pm$0.12 &  11.10$\pm$0.08 &  0.0403$\pm$0.0080 \\
NGC 5845 & ES,b       & n & -0.04$\pm$0.04 &  0.53$\pm$0.04 &  10.14$\pm$0.08 &  0.0247$\pm$0.0040 \\
NGC 5846 & E          &   & -0.07$\pm$0.04 & -0.13$\pm$0.08 &  11.18$\pm$0.09 & $<$0.0001 \\	 
NGC 6251 & E          &   & -0.04$\pm$0.04 &  1.05$\pm$0.04 &  11.51$\pm$0.08 &  1.0416$\pm$0.1047 \\ 
NGC 6861 & ES,b       & Y & -0.04$\pm$0.04 &  0.76$\pm$0.05 &  10.84$\pm$0.08 &  0.1063$\pm$0.0112 \\
NGC 6926 & S          &   &  0.19$\pm$0.03 &  3.79$\pm$0.04 &  11.08$\pm$0.08 & 19.2827$\pm$2.9770 \\ 
NGC 7052 & E          &   & -0.06$\pm$0.04 &  0.58$\pm$0.05 &  11.22$\pm$0.08 &  0.3386$\pm$0.0463 \\ 
NGC 7332 & S0         & N & -0.04$\pm$0.04 &  0.46$\pm$0.05 &  10.48$\pm$0.08 &  0.0497$\pm$0.0083 \\
NGC 7457 & S0         & N & -0.03$\pm$0.04 &  0.27$\pm$0.11 &  09.92$\pm$0.08 &  0.0150$\pm$0.0040 \\
NGC 7582 & S          &   &  0.59$\pm$0.03 &  3.29$\pm$0.04 &  10.59$\pm$0.08 &  8.9545$\pm$1.0156 \\ 
NGC 7619 & E          &   & -0.08$\pm$0.04 & -0.01$\pm$0.12 &  11.29$\pm$0.08 & $<$0.0003 \\	 
NGC 7768 & BCG (E)    &   & -0.10$\pm$0.04 & -0.38$\pm$0.05 &  11.44$\pm$0.09 & $<$0.0013 \\	 
UGC 3789 & S          &   &  0.09$\pm$0.03 &  3.22$\pm$0.04 &  10.51$\pm$0.08 &  2.0713$\pm$0.2073 \\

\hline
\end{tabular}

Column~2: Galaxy type \citep{Graham:Sahu:22a, Graham:Sahu:22b}.
Following \citet{Graham-S0}, NGC~4594 and NGC~2974 are considered S0 galaxies rather than S galaxies. 
The four ES,b galaxies noted there continue to be counted here among the S0 galaxies.
NGC~4395 and NGC~6926 are bulge-less S galaxies. 
Column~3: Dust `bin' for the S0 galaxies \citep[taken from ][]{Graham-S0}. 
Y = Strong Yes, 
y = weak yes,
n = nuclear dust,
N = No dust
(see Section~\ref{Sec_data}). 
Columns~4 \& 5: {\it WISE} colours. The W1, W2, W3 and W4 passbands are centred at 3.368, 4.618, 12.082 and 22.194
$\mu$m, respectively. 
Column~6: Logarithm of the galaxies' stellar mass derived from {\it WISE} photometry using a 
 \citet{2003PASP..115..763C} IMF.  (For Fig.~\ref{Fig1}, 
0.05 dex was subtracted from these values to create consistency with the 
\citet{2002Sci...295...82K} IMF.  For Fig.~\ref{Fig3}, no adjustment was made.
For the other figures, and the derivation of the sSFRs, the stellar masses 
from \citet{Graham:Sahu:22a} were used.) 
Column~7: Star formation rate in units of solar mass per year.

\end{table*}

%%%%%%%%%%%%%%%%%%%%%%%%%%%%%%%%%%%%%%%%%%%%%%%%%%
% Don't change these lines
\bsp    % typesetting comment
\label{lastpage}
\end{document}